\documentclass[fleqn,usenatbib]{mnras}

\usepackage[T1]{fontenc}
\usepackage{ae,aecompl}

\usepackage{graphicx}	
\usepackage{amsmath}	
\usepackage{amssymb}	

\newcommand{\rmf}{{\tt{RMF}}~}
\newcommand{\rprs}{$R_\mathrm{p}/R_\mathrm{s}$~}

\title[Three KELTs as seen by \textit{CHEOPS} and \textit{TESS}]{Rapidly rotating stars and their transiting planets: KELT-17b, KELT-19Ab, and KELT-21b in the \textit{CHEOPS} and \textit{TESS} era}

\author[Z. Garai et al.]{
Z. Garai,$^{1,2,3,4}$\thanks{E-mail: zgarai@gothard.hu}
T. Pribulla,$^{3}$
J. Kov\'{a}cs,$^{1,2,4}$
Gy. M. Szab\'{o},$^{1,2,4}$
A. Claret,$^{5,6}$
R. Kom\v{z}\'{i}k,$^{3}$
\newauthor
and E. Kundra$^{3}$
\\
$^{1}$MTA-ELTE Exoplanet Research Group, 9700 Szombathely, Szent Imre h. u. 112, Hungary\\
$^{2}$ELTE Gothard Astrophysical Observatory, 9700 Szombathely, Szent Imre h. u. 112, Hungary\\
$^{3}$Astronomical Institute, Slovak Academy of Sciences, 05960 Tatransk\'a Lomnica, Slovakia\\
$^{4}$MTA-ELTE Lend\"{u}let Milky Way Research Group, 9700 Szombathely, Szent Imre h. u. 112, Hungary\\ 
$^{5}$Instituto de Astrof\'{i}sica de Andaluc\'{i}a, CSIC, Apartado 3004, 18080 Granada, Spain\\
$^{6}$Dept. F\'{i}sica Te\'{o}rica y del Cosmos, Universidad de Granada, Campus de Fuentenueva s/n, 10871 Granada, Spain\\ 
}

\date{Accepted XXX. Received YYY; in original form ZZZ}

\pubyear{2015}

\begin{document}
\label{firstpage}
\pagerange{\pageref{firstpage}--\pageref{lastpage}}
\maketitle

\begin{abstract}
Rapidly rotating early-type main-sequence stars with transiting planets are interesting in many aspects. Unfortunately, several astrophysical effects in such systems are not well understood yet. Therefore, we performed a photometric mini-survey of three rapidly rotating stars with transiting planets, namely KELT-17b, KELT-19Ab, and KELT-21b, using the \textit{Characterising Exoplanets Satellite} (\textit{CHEOPS}), complemented with \textit{Transiting Exoplanet Survey Satellite} (\textit{TESS}) data, and spectroscopic data. We aimed at investigating the spin-orbit misalignment and its photometrical signs, therefore the high-quality light curves of the selected objects were tested for transit asymmetry, transit duration variations, and orbital precession. In addition, we performed transit time variation analyses, obtained new stellar parameters, and refined the system parameters. For KELT-17b and KELT-19Ab we obtained significantly smaller planet radius as found before. The gravity-darkening effect is very small compared to the precision of \textit{CHEOPS} data. We can report only on a tentative detection of the stellar inclination of KELT-21, which is about 60 deg. In KELT-17b and KELT-19Ab we were able to exclude long-term transit duration variations causing orbital precession. The shorter transit duration of KELT-19Ab compared to the discovery paper is probably a consequence of a smaller planet radius. KELT-21b is promising from this viewpoint, but further precise observations are needed. We did not find any convincing evidence for additional objects in the systems. 
   
\end{abstract}




\section{Introduction}
\label{intro}

\begin{figure*}
\centering
\centerline{
\includegraphics[width=55mm]{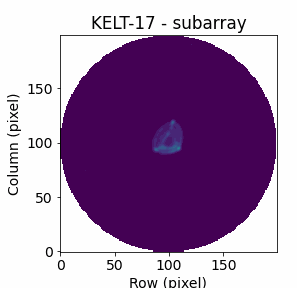}
\includegraphics[width=55mm]{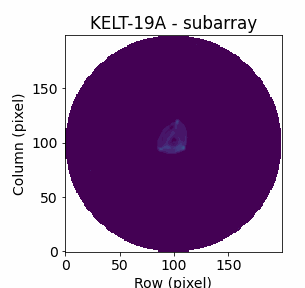}
\includegraphics[width=55mm]{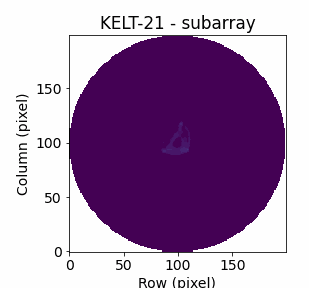}}
\caption{The $200 \times 200$ pixels subarrays of the target stars. The background stars were removed from the field of view. We can see the typical \textit{CHEOPS} shape of stars. The point spread function (PSF) is deliberately defocused to spread the flux over a number of pixels. In this way the telescope is less sensitive to inter-pixel variation and differences in flat fielding, just like other instruments. The distortion, i.e., the triangular shape, is due to the way in which the primary mirror is supported. For more details see \citet{Benz1}.}
\label{keltssubarray} 
\end{figure*}

Early-type main-sequence stars are rapidly rotating stars. Late-type stars (F5 and later) have deep convective envelopes and efficient magnetic dynamos, maintaining magnetic fields that transfer angular momentum to the stellar wind, thus slowing down the star's surface through magnetic braking. This is the so-called Kraft break \citep{Kraft1}. The rapid rotation at early-type stars leads to an oblate shape of the star and induces an equator-to-pole gradient in the effective temperature, called gravity darkening \citep{vonZeipel1, vonZeipel2}. The so-called von Zeipel theorem predicts that the flux emitted from the surface is proportional to the local effective gravity, thus the effect induces cooler temperatures at a rapidly rotating star's equator and hotter temperatures at the poles. However, the von Zeipel theorem is not strictly valid, hence it needs further investigation. \citet{Claret3} found important deviations from von Zeipel theorem in stars with envelopes in convective equilibrium. Moreover, \citet{Claret2} found significant deviations from von Zeipel theorem at the upper layers of a distorted star in radiative equilibrium. If an exoplanet transits a rapidly rotating star, distorted transit light curves are expected, as it was predicted by \citet{Barnes1}. If such asymmetries are measured, this can be used to determine the sky-projected angle $\lambda$ between the stellar rotational axis and the planet orbit normal, i.e., we can detect the spin-orbit misalignment. In addition, the stellar inclination $I_*$, which we define as the angle between the line of sight to the observer and the north pole of the star, can be derived and thus the true misalignment is possible to obtain. Another way to investigate the sky-projected spin-orbit misalignment is to measure the Rossiter-McLaughlin effect via radial velocities \citep{Rossiter1, McLaughlin1}. We can note, however, that radial velocity measurements in the case of early-type stars are difficult due to the rapid rotation. In the case of rapidly rotating stars the Doppler tomography in spectral line profiles, broadening function (BF) profiles, or least-squares deconvolution (LSD) profiles is applicable. With this technique the shadow of the transiting planet can also be detected, independently from the photometry \citep{Donati1}. Both techniques -- photometric and spectroscopic -- can be used to put constraints on the theories of planet formation and migration.

\citet{Szabo1} found an asymmetry in the rapidly rotating Kepler-13A transit light curve, consistent with the prediction of \citet{Barnes1}. Kepler-13A is the first known system, exhibiting a light-curve distortion due to spin-orbit misalignment and gravity darkening of a rapidly-rotating star. \citet{Barnes2} used this asymmetry to measure the sky projected spin-orbit misalignment angle $\lambda$. They determined $\lambda = 23 \pm 4$ deg, and $I_* = 138 \pm 4$ deg\footnote{This value reflects our definition of $I_*$, but \citet{Barnes2} originally presented this as follows: the north pole of the star is tilted away from the observer by $48 \pm 4$ deg.}, which gives the real misalignment of $56 \pm 4$ deg. This represents the first spin-orbit measurement obtained based on precise \textit{Kepler} photometry \citep{Borucki1, Borucki2}, although later \citet{Johnson3} obtained $\lambda = 58 \pm 2$ deg via Doppler tomography, which is a significantly different value for the sky projected spin-orbit misalignment angle. Moreover, the stellar rotation of Kepler-13A is in exact 5:3 resonance with the orbital period of the substellar companion, and the long-term transit duration variation with a rate of $(1.14 \pm 0.30) \times 10^{-6}$ d~cycle$^{-1}$ is due to the precession of its orbital plane \citep{Szabo2}. This long-term trend in the transit duration was confirmed using combined \textit{Kepler} and \textit{Transiting Exoplanet Survey Satellite} (\textit{TESS}) data \citep{Ricker1} by \citet{Szabo3}. We can say that the system Kepler-13A is a unique laboratory of many astrophysical effects.

\begin{table}
\centering
\caption{An overview of fundamental facts about the exoplanet host stars. Notes: H2000 = \citet{Hog1}, G2018 = \citet{Gaiadr2}.}
\label{stars}
\begin{tabular}{lcc}
\hline
\hline
Parameter [unit] & Value & Source\\ 
\hline
\hline
\multicolumn{3}{c}{KELT-17 (BD +14$^\circ$ 1881)}\\
RA [h:m:s] (J2000.0) & 08:22:28.2 & G2018\\
Dec [deg:m:s] (J2000.0) & +13:44:07.1 & G2018\\
$V$ [mag] & $9.23 \pm 0.02$ & H2000\\
$G$ [mag] & $9.2089 \pm 0.0003$ & G2018\\
\hline
\multicolumn{3}{c}{KELT-19A (BD +07$^\circ$ 1721)}\\
RA [h:m:s] (J2000.0) & 07:26:02.2 & G2018\\
Dec [deg:m:s] (J2000.0) & +07:36:56.8 & G2018\\
$V$ [mag] & $9.86 \pm 0.04$ & H2000\\
$G$ [mag] & $9.8633 \pm 0.0016$ & G2018\\
\hline
\multicolumn{3}{c}{KELT-21 (HD 332124)}\\
RA [h:m:s] (J2000.0) & 20:19:12.0 & G2018\\
Dec [deg:m:s] (J2000.0) & +32:34:51.7 & G2018\\
$V$ [mag] & $10.48 \pm 0.04$ & H2000\\
$G$ [mag] & $10.5415 \pm 0.0003$ & G2018\\
\hline
\hline
\end{tabular}
\end{table}

Recently, e.g., the asymmetric transit of the exoplanet KELT-9b was obtained and modeled using \textit{TESS} data \citep{Ahlers1}. KELT-9b is an ultra-hot Jupiter transiting a rapidly rotating early A-type star in a polar orbit \citep{Gaudi1}. This is the main reason, why KELT-9b is an interesting planet. The true spin-orbit misalignment angle was determined as $87 \pm 11$ deg by \citet{Ahlers1}. Another rapidly rotating A-type star with a transiting planet, i.e., WASP-189b, was observed also very recently using the \textit{Characterising Exoplanets Satellite} (\textit{CHEOPS}) space observatory \citep{Benz1}. From the asymmetric transit photometry of WASP-189b \citet{Lendl1} deduced the sky-projected spin-orbit misalignment angle of $\lambda = 86.4 \pm 4.4$ deg and the true misalignment of $85.4 \pm 4.3$ deg, in a good agreement with the previous measurement from spectroscopic observation \citep{Anderson1}. This result indicates that WASP-189b is in a polar orbit, similarly as KELT-9b. Since rapidly rotating early-type main-sequence stars with transiting planets can be similar to the "prototype" Kepler-13A system, in 2019 we proposed a mini-survey of such exoplanet hosts using the \textit{CHEOPS} space telescope, to constrain their planetary and stellar parameters, and to characterize the star-planet interactions in these cases. For this mini-survey we selected KELT-17, KELT-19A, and KELT-21 planetary systems. Fundamental facts about the exoplanet host stars are summarized in Table \ref{stars}. 

KELT-17b is a $1.3~M_\mathrm{Jup}$ and a $1.5~R_\mathrm{Jup}$ hot Jupiter, transiting the $V = 9.23$ mag main-sequence A-star KELT-17 in a 3.08-day misaligned orbit at $\lambda = 244.0$ deg, discovered by \citet{Zhou1}. The host star BD +14$^\circ$ 1881 ($M_* = 1.6~M_\odot$, $T_\mathrm{eff} = 7454$ K, $v \sin I_* = 44.2$ km~s$^{-1}$) is one of the most massive, hottest, and most rapidly rotating planet host stars. KELT-19Ab transits the $V = 9.86$ mag main-sequence A-star KELT-19A in a 4.61-day retrograde orbit ($\lambda = 180.3$ deg). The host star BD +07$^\circ$ 1721 is the first chemically peculiar Am-star, which hosts a hot Jupiter-type planet with a mass of $M_\mathrm{p} < 4.07~M_\mathrm{Jup}$ and a radius of about $1.9~R_\mathrm{Jup}$. Moreover, adaptive optics observations revealed a cooler stellar companion, KELT-19B, which is a G9V or K1V star. The stars have measured magnitude differences of $\Delta J = 2.50 \pm 0.06$ mag and $\Delta K_\mathrm{s} = 2.045 \pm 0.03$ mag \citep{Siverd1}. The projected separation is $0.64''$, so this system is a close analogue of the Kepler-13 system. KELT-21b is a $3.9~M_\mathrm{Jup}$ and a $1.5~R_\mathrm{Jup}$ hot Jupiter, transiting the $V = 10.5$ mag main-sequence A-star KELT-21 in a 3.6-day orbit, which is misaligned only slightly ($\lambda = 354.4$ deg). The host star HD 332124 has the highest projected rotational velocity among the exoplanet hosts ($v \sin I_* = 146$ km~s$^{-1}$), and it also appears to be somewhat metal poor \citep{Johnson1}. 

\begin{table*}
\centering
\caption{Log of \textit{CHEOPS} photometric observations of KELT-17b, KELT-19Ab, and KELT-21b transits (sorted by the targets and \textit{CHEOPS} visits). Table shows the time interval of individual observations, the applied exposure time, the number of obtained frames, the point-to-point root mean square ($RMS$) of the DRP-processed "OPTIMAL" light curves (see Sect. \ref{phot}), and the file key, which supports the fast identification of the observations in the \textit{CHEOPS} archive.}
\label{cheopsobslog}
\begin{tabular}{ccccccc}
\hline
\hline
Visit No. & Start date [UTC] & End date [UTC] & Exposure time [s] & Number of frames & $RMS$ [ppm] & File key\\
\hline
\hline
\multicolumn{7}{c}{KELT-17b}\\
1 & 2020-12-10 16:32 & 2020-12-11 01:16 & 55.1 & 330 & 420 & {\tt{CH\_PR210006\_TG000101}}\\
2 & 2020-12-16 20:55 & 2020-12-17 04:23 & 55.1 & 300 & 420 & {\tt{CH\_PR210006\_TG000102}}\\
3 & 2020-12-19 22:06 & 2020-12-20 06:12 & 55.1 & 292 & 380 & {\tt{CH\_PR210006\_TG000103}}\\
4 & 2021-02-16 10:33 & 2021-02-16 18:50 & 55.1 & 339 & 280 & {\tt{CH\_PR210006\_TG000104}}\\
\hline
\multicolumn{7}{c}{KELT-19Ab}\\
1 & 2020-11-27 12:31 & 2020-11-27 23:55 & 60.0 & 394 & 370 & {\tt{CH\_PR210006\_TG000201}}\\
2 & 2020-12-29 19:08 & 2020-12-30 07:06 & 60.0 & 460 & 520 & {\tt{CH\_PR210006\_TG000202}}\\
3 & 2021-01-26 11:21 & 2021-01-26 23:11 & 60.0 & 452 & 420 & {\tt{CH\_PR210006\_TG000203}}\\
4 & 2021-01-31 01:59 & 2021-01-31 13:33 & 60.0 & 461 & 440 & {\tt{CH\_PR210006\_TG000204}}\\
\hline
\multicolumn{7}{c}{KELT-21b}\\
1 & 2020-07-02 23:15 & 2020-07-03 08:32 & 60.0 & 333 & 720 & {\tt{CH\_PR210006\_TG000301}}\\
2 & 2020-07-21 01:04 & 2020-07-21 10:11 & 60.0 & 370 & 780 & {\tt{CH\_PR210006\_TG000302}}\\
3 & 2020-07-24 16:15 & 2020-07-25 01:34 & 60.0 & 379 & 720 & {\tt{CH\_PR210006\_TG000303}}\\
4 & 2020-08-26 03:24 & 2020-08-26 13:04 & 60.0 & 390 & 920 & {\tt{CH\_PR210006\_TG000304}}\\
\hline
\hline
\end{tabular}
\end{table*}

In this paper we aimed at refining the system parameters based on the obtained \textit{CHEOPS} photometry data, supplemented with several spectroscopic observations. Furthermore, based on the precise \textit{CHEOPS} transit light curves of the systems we aimed at searching for similar asymmetries, as it was detected in the Kepler-13A system and whether asymmetries are consistent with the prediction, coming from the previous spectroscopic results. Since the photometric follow-up observations of these systems using the \textit{CHEOPS} telescope can also reveal transit duration variations, and hence the orbital precession, our further scientific goal is to search for such indicators. Finally, transit time variations may also be detected, testing for additional planets in the systems, therefore we also included the search for such variations in our scientific aims. The paper is organized as follows. In Section \ref{obs} a brief description of instrumentation and data reduction is given. We summarize the spectral analysis and the derived stellar parameters in Section \ref{spec}. The fundamental analysis of the \textit{CHEOPS} transits and the obtained system parameters are described and discussed in Section \ref{sysparamsfromcheops}. In Section \ref{spinorb} we analyze the \textit{CHEOPS} light curves from the viewpoint of spin-orbit misalignment. Search for long-term transit duration variations and transit time variations is detailed in Section \ref{tdvttv}. We summarize our findings in Section \ref{concl}.                                  

\section{Observations and data reduction}
\label{obs}

\subsection{Transit photometry}
\label{phot}

The transits of KELT-17b, KELT-19Ab, and KELT-21b were observed photometrically using the \textit{CHEOPS} space observatory \citep{Benz1}. This is the first European space mission dedicated primarily to the study of known exoplanets. It consists of a 32 cm mirror diameter telescope\footnote{The primary mirror is partly blocked by the secondary mirror spider legs and has a central cut-out to allow the beam to pass through on to the CCD. This means that the effective mirror diameter of \textit{CHEOPS} is 30 cm.} based on a Ritchey-Chr\'etien design. The photometric detector is a single CCD camera covering the wavelength range from 330 to 1100 nm with a field of view of $0.32~\mathrm{deg}^2$. The payload design and operation have been optimized to achieve ultra-high photometric stability, achieving a photometric precision of 20 ppm on observations of a G5-type star in 6 hours, and 85 ppm observations of a K5-type star in 3 hours. 20\% of the science time on \textit{CHEOPS} is available to the astronomical community through a Guest Observers Programme that is open to the science community as a whole. \textit{CHEOPS} observations used in this work were obtained within the first cycle of the Guest Observers Programme, proposal ID 006, entitled "Rapidly rotating stars and their transiting planets: a unique laboratory of many astrophysical effects"\footnote{See the list of approved programs at \url{https://www.cosmos.esa.int/web/cheops-guest-observers-programme/ao-1-programmes}.} (PI: Z. Garai).

Based on the literature transit duration values \citep{Zhou1, Siverd1, Johnson1}, at least 0.1-long phase interval is needed at every transit event (phases $0.95-1.05$ around the mid-transit time) to properly cover the transit and the neighbor out-of-transit phases with observations. Therefore, we proposed 5 orbits\footnote{\textit{CHEOPS} revolves around the Earth in Sun-synchronous, low-Earth  orbit (700 km altitude). The spacecraft completes one orbit around the Earth in 99 min.} per visit\footnote{A visit is a sequence of successive \textit{CHEOPS} orbits devoted to observing a given target.} and 4 visits in the case of KELT-17b, 7 orbits per visit and 4 visits in the case of KELT-19Ab, and 6 orbits per visit and 4 visits in the case of KELT-21b, 72 orbits in total, including interruptions\footnote{Interruptions happen when the target is hidden by the Earth during an Earth occultation, or barely visible due to stray light from the illuminated Earth limb or particle hits during passages through the South Atlantic anomaly.}. Due to the interruptions several phase gaps occurred during a single visit. The predicted observing efficiencies\footnote{The observing efficiency is the ratio between the amount of science observing time available during a visit (excluding the interruptions) and the total amount of time in a visit (including the interruptions).} were 59\%, 58\%, and 63\% in the cases of KELT-17b, KELT-19Ab, and KELT-21b, respectively. Further details about the \textit{CHEOPS} observations can be found in Table \ref{cheopsobslog}.        

From the \textit{CHEOPS} detector, which has $1024 \times 1024$ pixels, a $200 \times 200$ pixels subarray is extracted around the target point spread function, which is used to compute the photometry (see Fig. \ref{keltssubarray}). The \textit{CHEOPS} Data Reduction Pipeline -- DRP \citep{Hoyer1} provides aperture photometry of these subarray frames. It performs several image corrections, including bias-, dark-, and flat-corrections, contamination estimation and background-star correction. The DRP produces 4 different light curve types for each visit: "DEFAULT" -- estimated using the default aperture radius of 25 pixels, "OPTIMAL" -- the aperture radius is automatically set based on the signal-to-noise ratio, "RINF" -- using the aperture radius of $0.9 \times 25$ pixels, and "RSUP" -- using the aperture radius of $1.2 \times 25$ pixels. According to the relatively long exposure times (55.1 and 60.0 s), the data were transferred to the Earth without stacking the individual images together, which means that the so-called imagettes with a smaller radius of 30 pixels are not available in these cases. 

The DRP-processed \textit{CHEOPS} light curves were downloaded from the CHEOPS Archive Browser\footnote{See \url{https://cheops-archive.astro.unige.ch/archive_browser/}.}. We first ran several modeling tests using the {\tt{RMF}} code, described in Sect. \ref{spinorb}, in order to select the best light-curve type offered by the archive. Since there is no significant difference among the light-curve types from the viewpoint of precision, we decided to use the "OPTIMAL" light curves during our analysis procedure. 

\subsection{Target spectroscopy}
\label{spectra}

Besides the transit photometry, spectra of the exoplanet hosts were also recorded several times to characterize these stars. The spectroscopic observations were obtained at the Skalnat\'{e} Pleso Observatory (Slovakia), using the 1.3 m f/8.36 Astelco Alt-azimuthal Nasmyth-Cassegrain reflecting telescope, equipped with a fiber-fed echelle spectrograph of MUSICOS design \citep{Baudrand1}. Its fiber injection and guiding unit (FIGU) is mounted in the Nasmyth focus of the telescope. The FIGU is connected to the calibration unit (ThAr hollow cathode lamp, tungsten lamp, blue LED) in the control room and to the echelle spectrograph itself in the room below the dome, where the temperature is stable. The spectra were recorded by an Andor iKon-936 DZH $2048 \times 2048$ pixels CCD camera. The spectral range of the instrument is 4250 -- 7375 \AA~ in 56 echelle orders. The maximum resolution of the spectrograph reaches $R \approx 38~000$ around 6000 \AA. The exposure time was 900 s in all cases. Three raw spectra were obtained consecutively during an observing night. More details about the spectroscopic observations can be found in the observations log (see Table \ref{spectroobslog}).       

\begin{table}
\centering
\caption{Log of spectroscopic observations of KELT-17, KELT-19A, and KELT-21 (sorted by the targets). Table shows the time interval of observations and the signal-to-noise ratio ($S/N$) of the combined spectra at 5500 \AA. The $S/N$ was calculated as $S/N = \sqrt{(S/N)_1{^2} + (S/N)_2{^2} + (S/N)_3{^2}}$, where $(S/N)_\mathrm{n}$ is the signal-to-noise ratio of individual spectra.}
\label{spectroobslog}
\begin{tabular}{ccc}
\hline
\hline
Start date [UTC] & End date [UTC] & $S/N$\\
\hline
\hline
\multicolumn{3}{c}{KELT-17}\\
2020-02-20 21:40:23 & 2020-02-20 22:28:27 & 46.5\\
2020-03-05 19:20:11 & 2020-03-05 20:08:10 & 30.3\\
2020-03-17 20:39:25 & 2020-03-17 21:27:24 & 42.8\\
2020-11-25 03:19:07 & 2020-11-25 04:07:07 & 38.0\\
2021-01-11 02:41:40 & 2021-01-11 03:29:40 & 38.0\\
2021-01-31 21:49:15 & 2021-01-31 22:37:16 & 48.7\\
2021-02-20 22:11:30 & 2021-02-20 22:59:29 & 44.6\\
2021-02-22 21:09:22 & 2021-02-22 21:57:22 & 40.5\\
\hline
\multicolumn{3}{c}{KELT-19A}\\
2020-02-08 21:17:29 & 2020-02-08 22:05:29 & 38.5\\
2020-02-20 19:44:45 & 2020-02-20 20:33:33 & 30.9\\
2020-03-17 19:06:11 & 2020-03-17 19:54:11 & 31.3\\
2020-12-02 01:23:02 & 2020-12-02 02:11:02 & 38.3\\
2020-12-13 00:35:12 & 2020-12-13 01:23:12 & 40.2\\
2020-12-19 23:42:37 & 2020-12-20 00:30:37 & 44.5\\
2021-01-10 23:54:41 & 2021-01-11 00:42:40 & 30.0\\
2021-02-21 18:44:05 & 2021-02-21 19:32:06 & 37.5\\
2021-02-22 19:33:07 & 2021-02-22 20:21:08 & 29.9\\
\hline
\multicolumn{3}{c}{KELT-21}\\
2020-07-05 22:30:27 & 2020-07-05 23:18:28 & 25.8\\
2020-07-30 22:40:10 & 2020-07-30 23:28:05 & 17.3\\
2020-08-01 22:32:39 & 2020-08-01 23:15:38 & 23.2\\
2020-08-12 22:56:31 & 2020-08-12 23:44:32 & 27.7\\
2020-09-14 20:33:48 & 2020-09-14 21:21:48 & 24.3\\
2020-10-28 16:31:23 & 2020-10-28 17:19:23 & 22.5\\
2020-11-09 17:46:08 & 2020-11-09 18:36:14 & 29.1\\
2020-11-10 17:04:23 & 2020-11-10 17:52:24 & 27.6\\
\hline
\hline
\end{tabular}
\end{table}

The raw spectra were reduced using {\tt{IRAF}} package tasks, {\tt{Linux}} shell scripts, and {\tt{FORTRAN}} programs similarly, as it was described in \citet{Pribulla1} and in \citet{Garai1}. In the first step, master dark frames were produced. In the second step, the photometric calibration of the frames was done using dark and flat-field frames. Bad pixels were cleaned using a bad pixel mask, and cosmic hits were removed using the program of \citet{Pych1}. Order positions were defined by fitting Chebyshev polynomials to tungsten lamp and blue LED spectrum. In the following step, scattered light was modeled and subtracted. Aperture spectra were then extracted for both object and ThAr frames, and then the resulting 2D spectra were dispersion-solved. Two-dimensional spectra were finally combined to 1D spectra rebinned to 4250 -- 7375 \AA~ wavelength range with a 0.05 \AA~ step, i.e. about 2 -- 4 times the spectral resolution. 

The obtained 1D spectra were combined to increase the signal-to-noise ratio using {\tt{iSpec}}\footnote{See \url{https://www.blancocuaresma.com/s/iSpec}.} \citep{Blanco1, Blanco2}. As first, three consecutive spectra of the same night were combined with the assumption that there is no substantial difference (Doppler shift) between them. For the following steps we used these averaged (median) spectra. We shifted all of the spectra (KELT-17 -- 8 spectra, KELT-19A -- 9 spectra, KELT-21 -- 8 spectra) into the rest frame combined the barycentric correction and the intrinsic radial velocity correction into one step applying the {\tt{iSpec}} cross-correlation routine. We cross-correlated the spectra with a template from the \citet{Munari1} synthetic spectrum library. According to the literature values for the stellar parameters, i.e., the effective temperature, the surface gravity, and metallicity \citep{Zhou1, Siverd1, Johnson1}, all three host stars are very similar, therefore we selected the template file from the spectrum library, which corresponds to $T_\mathrm{eff} = 7500$ K, $\log g = 4.0$ cgs and [M/H] = -0.5 dex. After shifting the spectra into the rest frame we averaged them via median into a final spectrum per object with setting the resolution to $R = 20~000$ (average resolution of the spectrograph), and setting the sampling to 0.05 \AA. We then corrected for the depression between 4800 and 5540 \AA~ (KELT-17 and KELT-19A), and slightly shifted the overall continuum level upward with a value of about 0.02 to set it to be 1.0 as much as possible. In this way we obtained the final averaged spectra for the host stars, which we further analyzed to obtain basic stellar parameters (see Sect. \ref{spec}).  

\section{Stellar parameters from spectra}
\label{spec}

\subsection{Data analysis}

\begin{figure*}
\centering
\centerline{
\includegraphics[width=55mm]{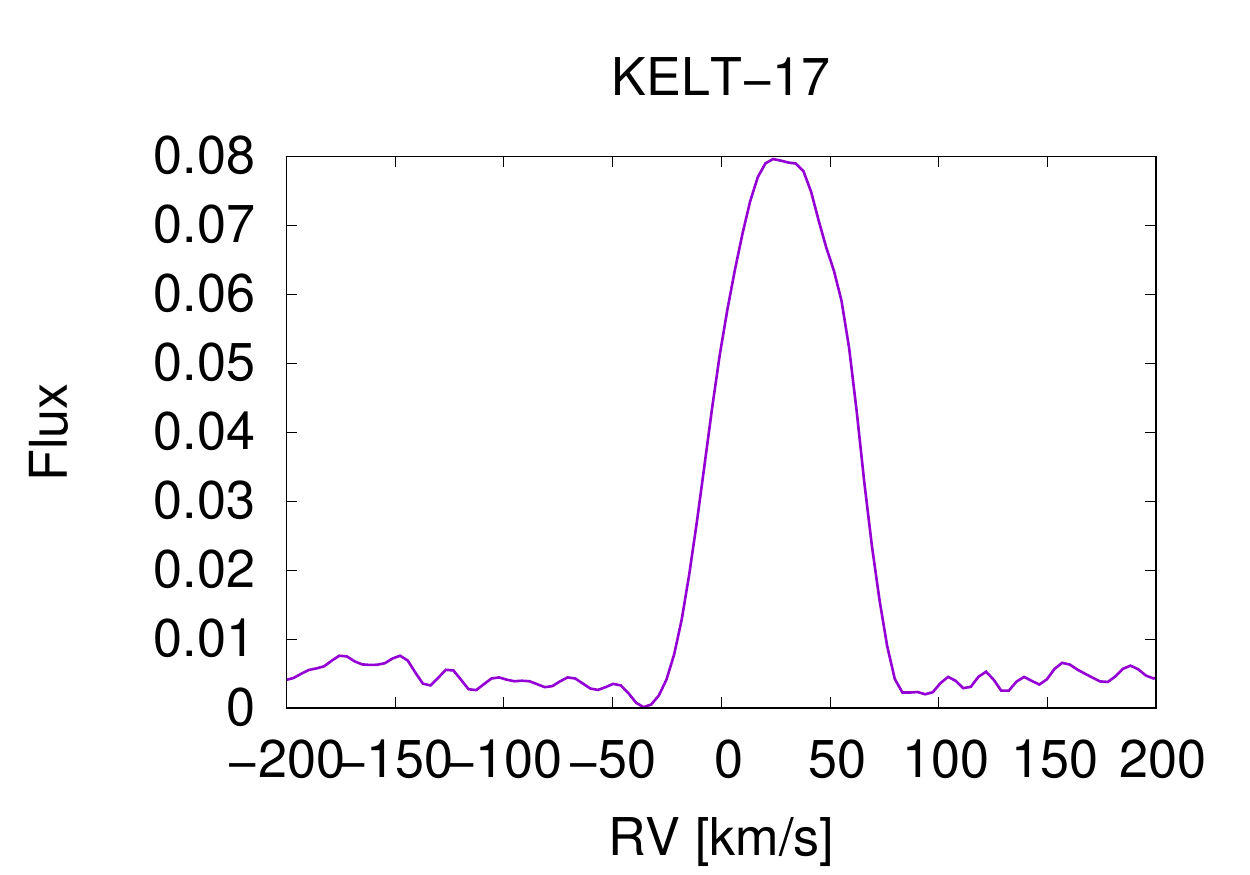}
\includegraphics[width=55mm]{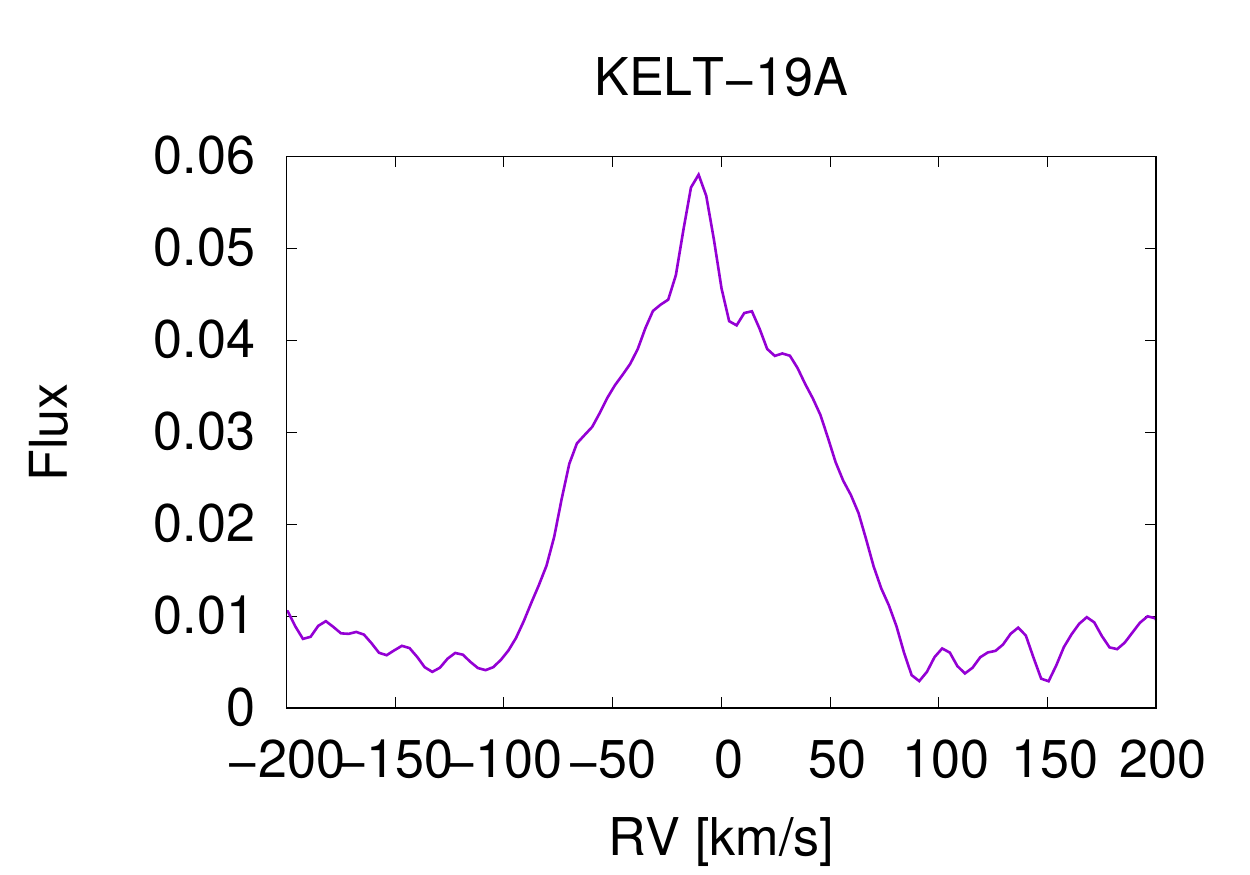}
\includegraphics[width=55mm]{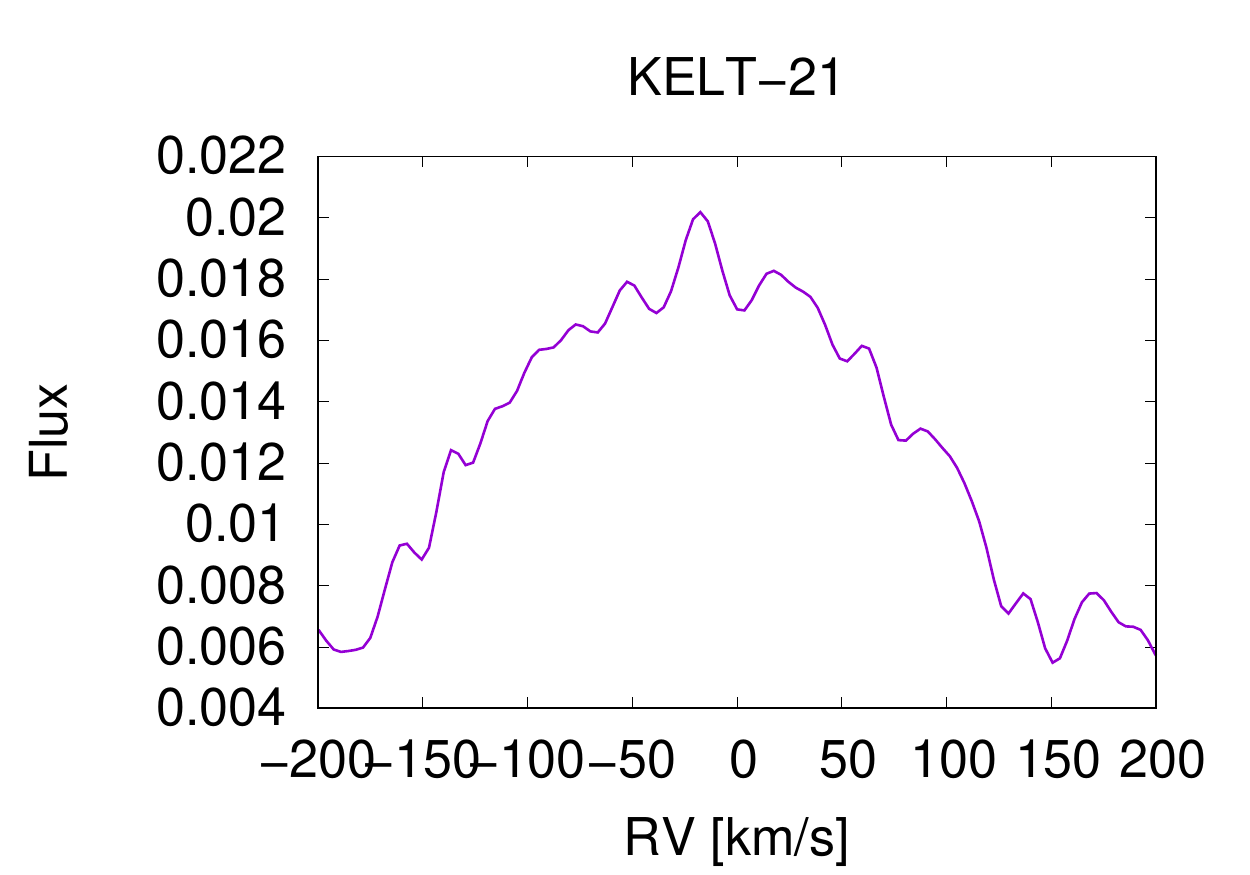}}
\caption{Selected broadening functions of the host stars, smoothed differently, according to the rotational velocities. The radial velocities (RV) are barycentric. The secondary component KELT-19B is well visible as a narrow peak on top of the broad profile of the primary component KELT-19A.}
\label{broadfunct}
\end{figure*}

\begin{figure*}
\centering
\centerline{
\includegraphics[width=\columnwidth]{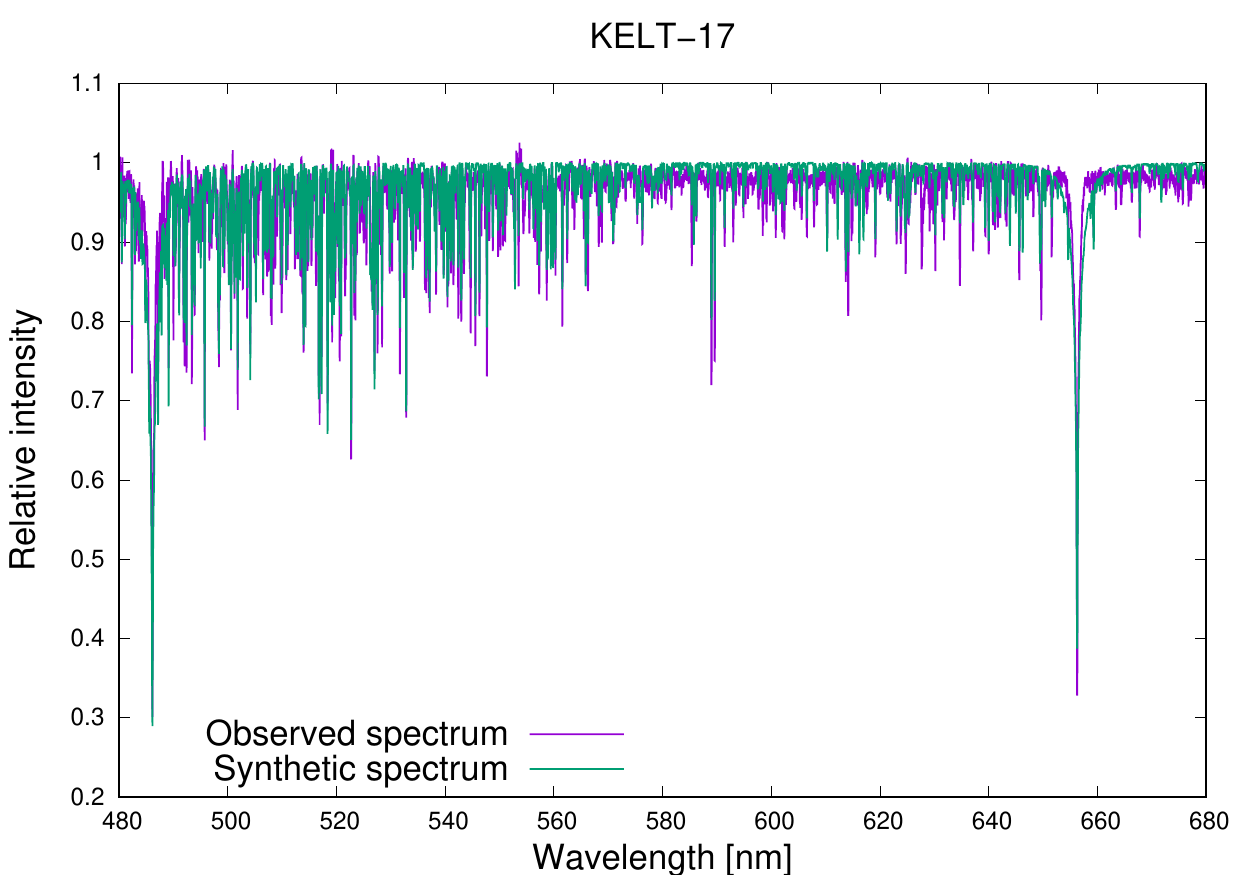}
\includegraphics[width=\columnwidth]{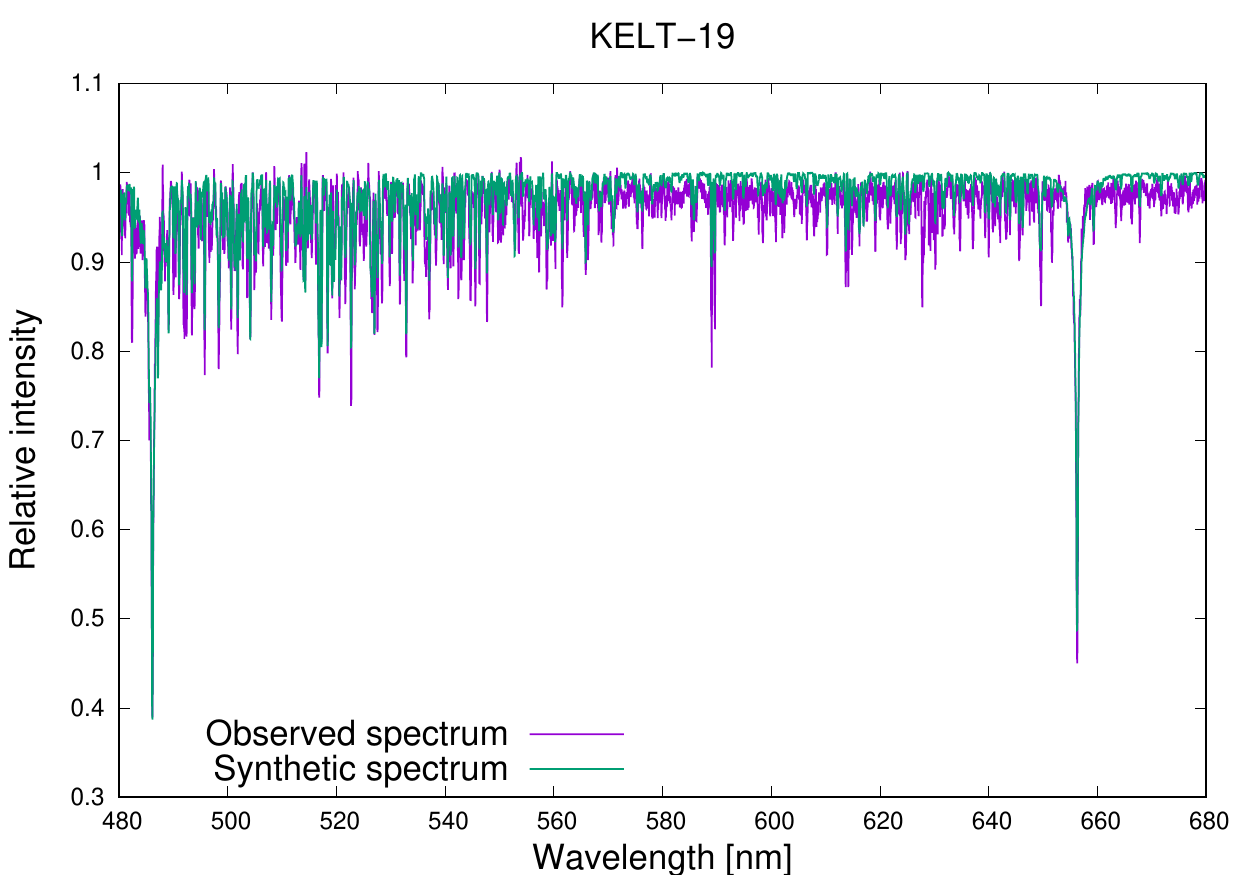}}
\caption{The final averaged spectrum of the host stars KELT-17 (left-hand panel) and KELT-19A (right-hand panel), overplotted with the synthetic spectrum.}
\label{kelt1719spectrum} 
\end{figure*} 

\begin{figure}
\includegraphics[width=\columnwidth]{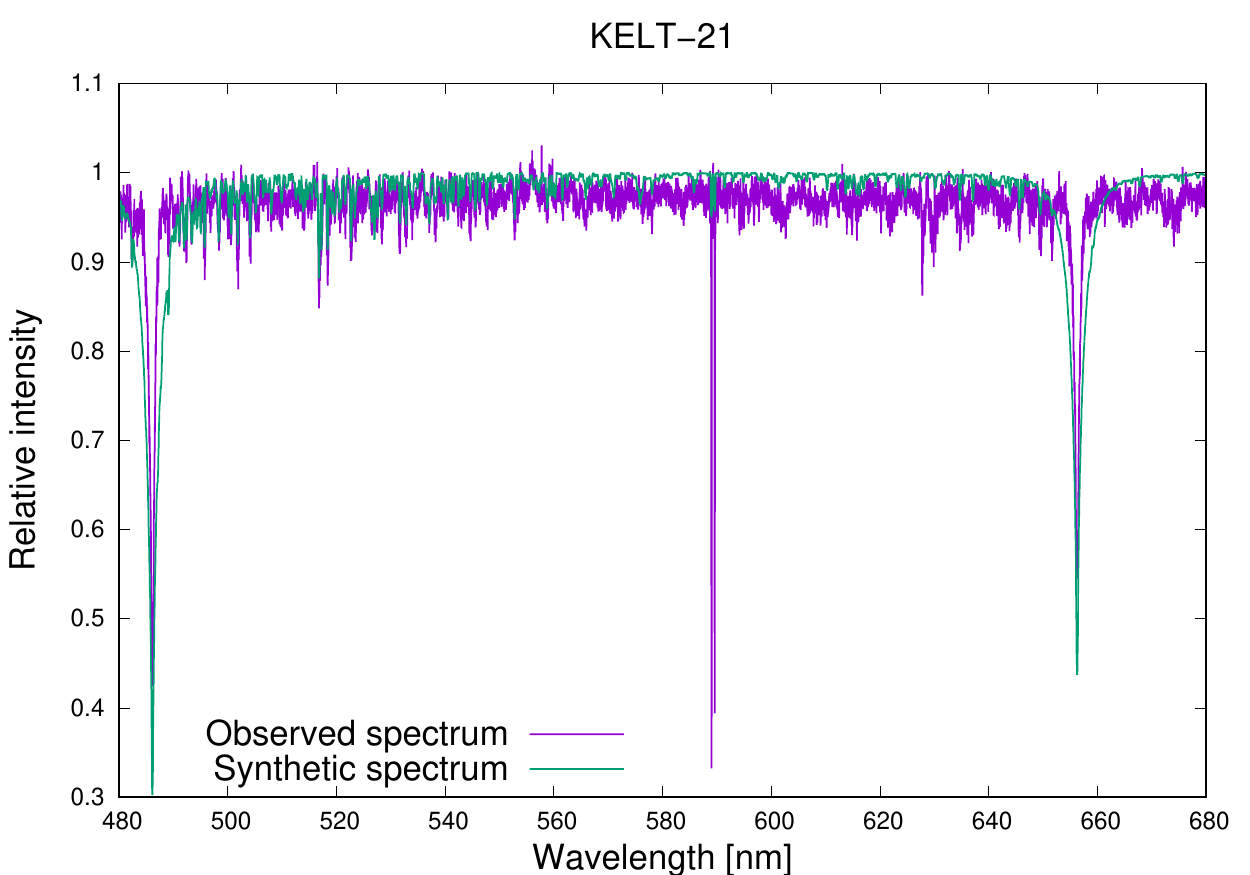}
\caption{As in Fig. \ref{kelt1719spectrum}, but for KELT-21.}
\label{kelt21spectrum} 
\end{figure} 

We tried to fit the final averaged spectra with several spectral synthesis softwares, but only the {\tt{FASMA}}\footnote{See \url{http://www.iastro.pt/fasma/}.} code \citep{Andreasen1, Tsantaki1} led to reasonable results. This could be due to the relatively high temperature of the host stars. The spectral synthesis softwares, e.g., \texttt{SPECTRUM} \citep{Gray1}, \texttt{iSpec} \citep{Blanco1, Blanco2}, or \texttt{SME} \citep{Valenti1}, work more effectively for the spectra with $T_\mathrm{eff}$ between 5000 and 6000 K. The spectroscopic analysis in the \texttt{FASMA} software is based on the spectral synthesis technique using the radiative transfer code {\tt{MOOG}}\footnote{See \url{https://www.as.utexas.edu/~chris/moog.html}.}. The model atmospheres are generated by the ATLAS9 program\footnote{See \url{http://research.iac.es/proyecto/ATLAS-APOGEE/}.} \citep{Meszaros1}, and all grids are based on 1D atmosphere in LTE. {\tt{FASMA}} includes the parameter optimization procedure based on the Levenberg-Marquardt algorithm. Uncertainties on fitted parameters are estimated applying the covariance matrix method, see \citet{Gonzalez2} and \citet{Gonzalez1}.

The software offers the option to the user either to provide initial guesses for the parameters or set the spectral type and luminosity class of the star. We used the first option and set the starting parameters as follows. Since the host stars are very similar, we uniformly set the effective temperature ($T_\mathrm{eff} = 7500$ K), the surface gravity ($\log g = 4.0$ cgs), metallicity ([Fe/H] = 0.0 dex), and the stellar microturbulent velocity ($v_\mathrm{mic} = 1.0$ km~s$^{-1}$). The stellar macroturbulent velocity was fixed during the analysis to the value of $v_\mathrm{mac} = 0.0$ km~s$^{-1}$, which is justified by the radiative envelopes of A-type stars and by the definition of the macroturbulent velocity in \texttt{FASMA}, which describes the motion in larger atmospheric cells \citep{Tsantaki1}. We did not adjust the projected rotational velocity during the analysis, as well. This parameter was derived based on the BF technique \citep{Rucinski1} and then we adopted and fixed. Examples of BFs are depicted in the panels of Fig. \ref{broadfunct}. For KELT-17 we obtained $v \sin I_* = 48.49 \pm 0.15$ km~s$^{-1}$, in the case of KELT-19A we got $v \sin I_* = 86.36 \pm 0.21$ km~s$^{-1}$, and for KELT-21 we derived $v \sin I_* = 141.9  \pm 2.4$ km~s$^{-1}$. In the latter case we used 99 km~s$^{-1}$ in the software, because this is the possible upper limit in \texttt{FASMA}. If $v \sin I_*$ is set to zero at the start of the fitting process, it will increase systematically at each step until it reaches the numerical limit of the program, so it would probably be closer to the true value of about 141 km~s$^{-1}$ if the program did not have this limit. We tested the effect of this constraint on the fitted parameters during calculations, where the $v \sin I_*$ parameter was fixed to the values of 50, 60, 70, 80, 90, and 99 km~s$^{-1}$. We found that $v \sin I_* > 60$ km~s$^{-1}$ has negligible effect on the fitted parameters. Finally, we can note that the procedure is relatively independent from the initial conditions and the starting parameters affect only the computing time of the fitting procedure.

\begin{table}
\centering
\caption{An overview of the stellar parameters obtained from the spectra of KELT-17, KELT-19A and KELT-21, compared to the previously published parameters.}
\label{spectrares}
\begin{tabular}{lll}
\hline
\hline
Parameter [unit] & This work & \citet{Zhou1}\\
\hline
\hline
\multicolumn{3}{c}{KELT-17}\\
$M_*$ [$M_\odot$] & -- & $1.635 \pm 0.066$\\
$R_*$ [$R_\odot$] & -- & $1.645 \pm 0.060$\\
$I_*$ [deg] & -- & $94 \pm 10$\\
$T_\mathrm{eff}$ [K] & $7109 \pm 252$ & $7454 \pm 49$\\
$\log g$ [cgs] & $4.28 \pm 0.39$ & $4.220 \pm 0.024$\\
Fe/H [dex] & $-0.08 \pm 0.12$ & $-0.018 \pm 0.074$\\
$v_\mathrm{mic}$ [km~s$^{-1}$] & $3.31 \pm 0.35$ & --\\
$v \sin I_*$ [km~s$^{-1}$] & $48.49 \pm 0.15$ & $44.2 \pm 1.5$\\
\hline
\hline
Parameter [unit] & This work & \citet{Siverd1}\\
\hline
\hline
\multicolumn{3}{c}{KELT-19A}\\
$M_*$ [$M_\odot$] & -- & $1.62 \pm 0.25$\\
$R_*$ [$R_\odot$] & -- & $1.830 \pm 0.099$\\
$T_\mathrm{eff}$ [K] & $6643 \pm 391$ & $7500 \pm 110$\\
$\log g$ [cgs] & $3.56 \pm 0.63$ & $4.127 \pm 0.029$\\
Fe/H [dex] & $-0.38 \pm 0.21$ & $-0.12 \pm 0.51$\\
$v_\mathrm{mic}$ [km~s$^{-1}$] & $2.62 \pm 0.42$ & --\\
$v \sin I_*$ [km~s$^{-1}$] & $86.36 \pm 0.21$ & $84.8 \pm 2.0$\\
\hline
\hline
Parameter [unit] & This work & \citet{Johnson1}\\
\hline
\hline
\multicolumn{3}{c}{KELT-21}\\
$M_*$ [$M_\odot$] & -- & $1.458 \pm 0.029$\\
$R_*$ [$R_\odot$] & -- & $1.638 \pm 0.034$\\
$T_\mathrm{eff}$ [K] & $8210 \pm 771$ & $7598 \pm 84$\\
$\log g$ [cgs] & $4.53 \pm 1.12$ & $4.173 \pm 0.015$\\
Fe/H [dex] & $-0.19 \pm 0.36$ & $-0.405 \pm 0.033$\\
$v_\mathrm{mic}$ [km~s$^{-1}$] & $0.68 \pm 1.08$ & --\\
$v \sin I_*$ [km~s$^{-1}$] & $141.9 \pm 2.4$ & $146.03 \pm 0.48$\\
\hline
\hline
\end{tabular}
\end{table}

\subsection{Results of target spectroscopy}

The obtained parameters are summarized and compared to the previously published parameters in Table \ref{spectrares}. We briefly discuss these parameters in the following subsections. The observed and averaged stellar spectra, overplotted with the synthetic spectra are depicted in the Figs. \ref{kelt1719spectrum} and \ref{kelt21spectrum}. 

\subsubsection{KELT-17}               

KELT-17 has a mass of $1.635 \pm 0.066~M_\odot$ and a radius of $1.645 \pm 0.060~R_\odot$ \citep{Zhou1}. Based on an independent differential rotation analysis, presented by \citet{Zhou1}, there is some information about the stellar inclination of KELT-17. They found $I_* = 94 \pm 10$ deg, which means that the star is seen nearly equator-on. Using the {\tt{FASMA}} code we obtained the stellar parameters of $T_\mathrm{eff} = 7109 \pm 252$ K, $\log g = 4.28 \pm 0.39$ cgs, [Fe/H] = $-0.08 \pm 0.12$ dex, and $v_\mathrm{mic} = 3.31 \pm 0.35$ km~s$^{-1}$, which are in a $3\sigma$ agreement with the previously derived stellar parameters, obtained by \citet{Zhou1}.

\subsubsection{KELT-19A}         

As noted in Sect. \ref{intro}, KELT-19 is a visual double star. The host star is the primary component KELT-19A, which has peculiar abundance pattern that is indicative of it belonging to the class of metallic-line mean sequence Am stars \citep{Siverd1}. It has a mass of $1.62 \pm 0.25~M_\odot$ and a radius of $1.830 \pm 0.099~R_\odot$ as derived by the discoverers. Since the star rotates faster than KELT-17, the parameters are determined less precisely than in the previous case. Moreover, spectrum of KELT-19A includes light contamination from the companion star KELT-19B, which is a cooler G-, or K-type mean sequence star. This means a possible systematic bias, even if the secondary component contributes with a very low signal to the composite spectrum. The light contribution of the companion depends on the seeing conditions. Its parameters are quite uncertain, thus difficult to disentangle the spectrum. These conditions allowed us to obtain the following stellar parameters with {\tt{FASMA}}: $T_\mathrm{eff} = 6643 \pm 391$ K, $\log g = 3.56 \pm 0.63$ cgs, [Fe/H] = $-0.38 \pm 0.21$ dex, and $v_\mathrm{mic} = 2.62 \pm 0.42$ km~s$^{-1}$. The parameters are in a $3\sigma$ agreement to those of derived by \citet{Siverd1}.

\subsubsection{KELT-21}

KELT-21 was also analyzed using the {\tt{FASMA}} software. This is the most rapid rotator in our sample, therefore several stellar parameters are determined poorly, or with a more than $3\sigma$ difference in comparison with the previously derived parameters, obtained by \citet{Johnson1}. KELT-21 is metal poor, which is unusual for relatively young hot stars \citep{Johnson1}. Adaptive optics imaging reveal two likely companions of KELT-21 in the projected distance of about $1.2''$, but the possible KELT-21B and KELT-21C companions are much fainter than KELT-21. The contrast is about 7.3 mag, hence they should have a negligible effect on the spectrum of the host star. The main problem at this star, which affects the stellar synthesis, is the rapid rotation with $v \sin I_* = 141.9 \pm 2.4$ km~s$^{-1}$. Other derived stellar parameters are $T_\mathrm{eff} = 8210 \pm 771$ K, $\log g = 4.53 \pm 1.12$ cgs, [Fe/H] = $-0.19 \pm 0.36$ dex, and $v_\mathrm{mic} = 0.68 \pm 1.08$ km~s$^{-1}$. The mass of the star is $1.458 \pm 0.029~M_\odot$ and its radius is $1.638 \pm 0.034~R_\odot$ \citep{Johnson1}.       

\section{System parameters from \textit{CHEOPS} transits}
\label{sysparamsfromcheops}

\subsection{Individual transit analysis}
\label{indanalysis}

\begin{figure*}
\centering
\centerline{
\includegraphics[width=\columnwidth]{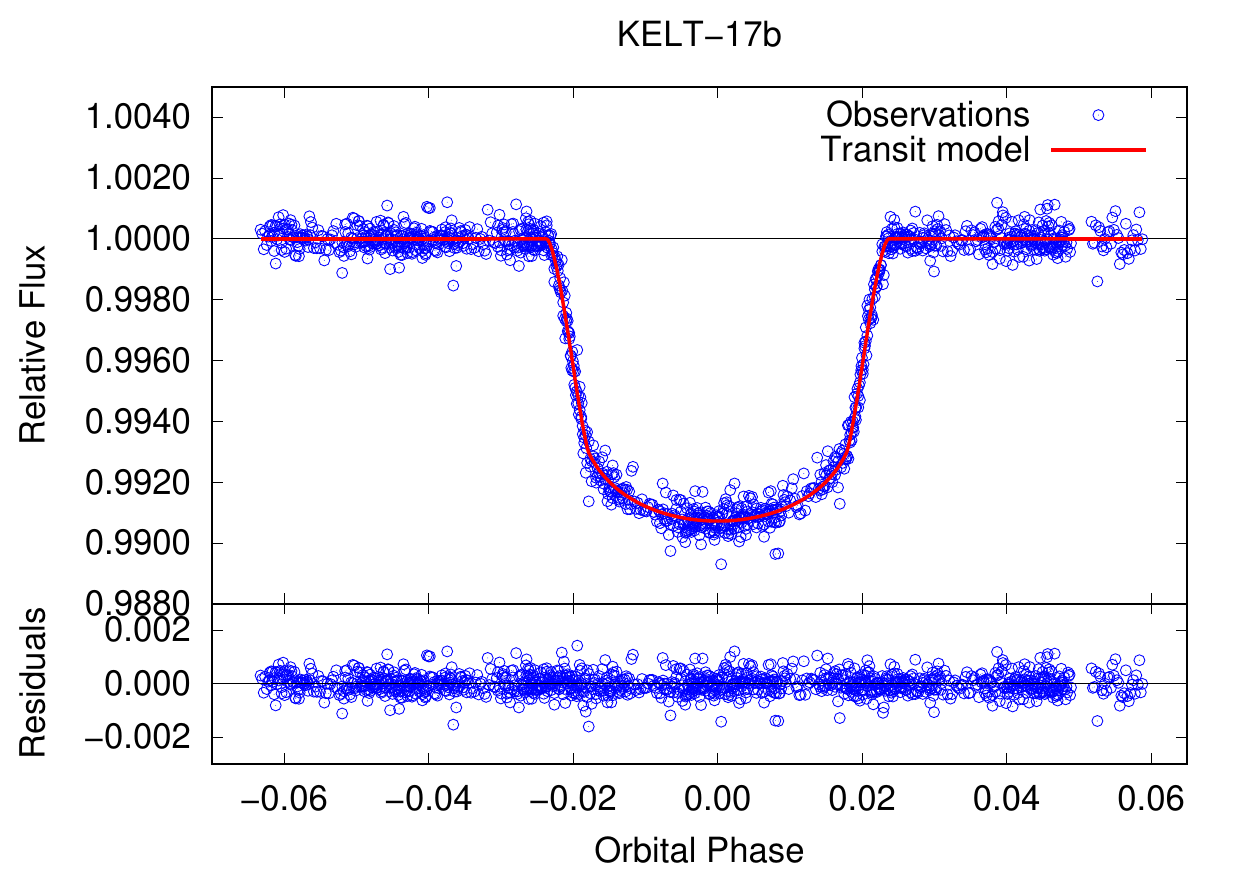}
\includegraphics[width=\columnwidth]{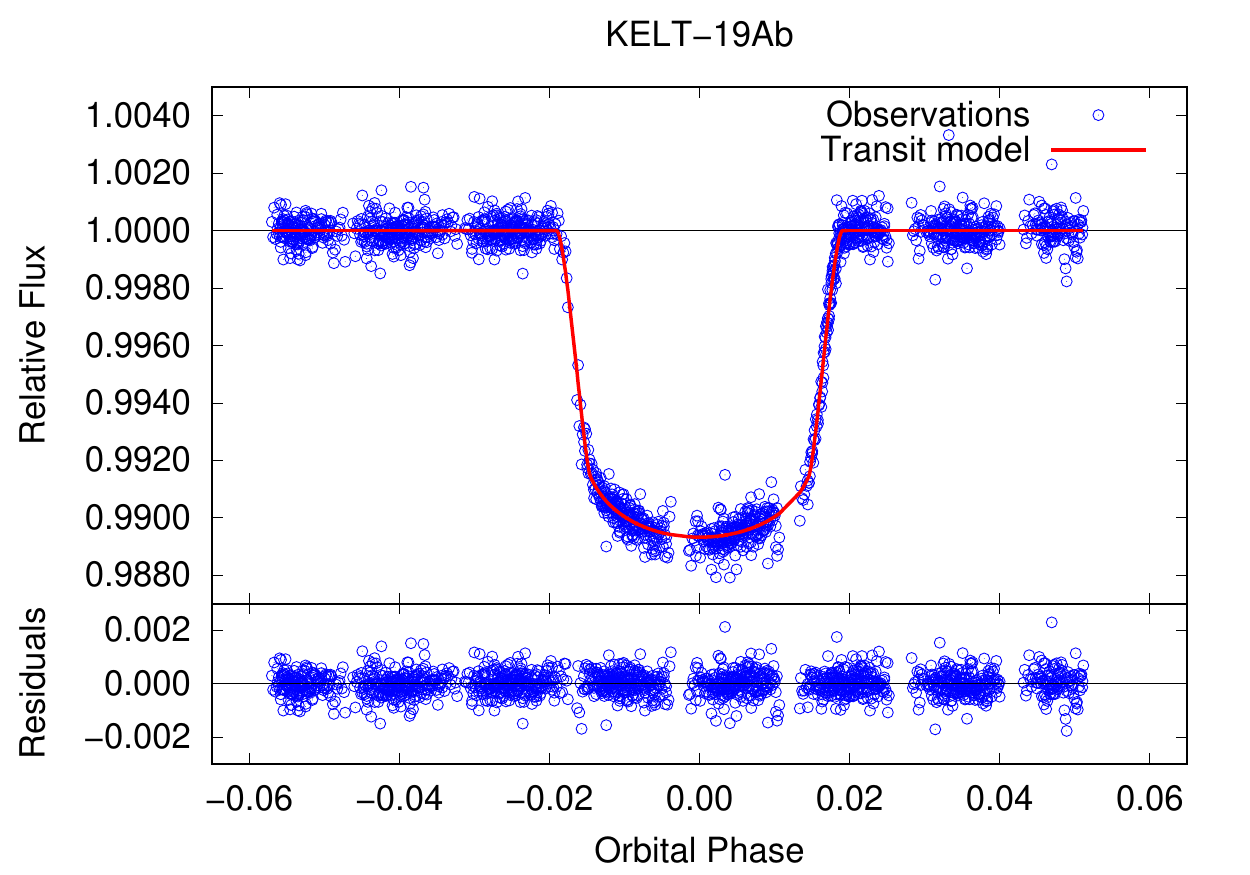}}
\caption{Phase-folded \textit{CHEOPS} transit light curves of KELT-17b (left-hand panel) and KELT-19Ab (right-hand panel), overplotted with the best-fitting \texttt{pycheops} models. Residuals are also shown (bottom panels). During the joint modeling procedure all individual \textit{CHEOPS} light curves were combined and fitted simultaneously.}
\label{kelt1719fittedlc} 
\end{figure*} 

\begin{figure}
\includegraphics[width=\columnwidth]{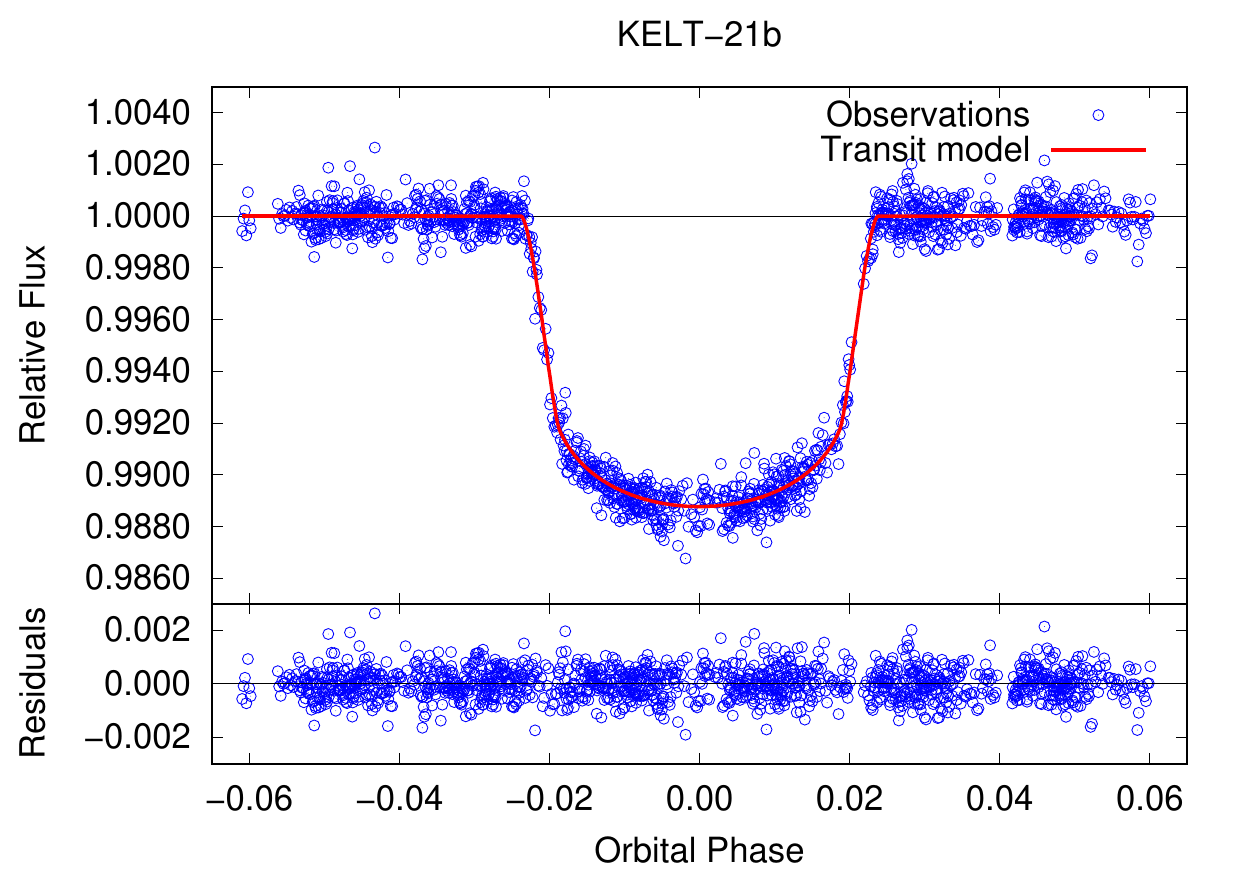}
\caption{As in Fig. \ref{kelt1719fittedlc}, but for KELT-21b.}
\label{kelt21fittedlc} 
\end{figure} 

\begin{table*}
\centering
\caption{An overview of the \texttt{pycheops} best-fitting and derived parameters of the exoplanets KELT-17b, KELT-19Ab, and KELT-21b obtained from the \textit{CHEOPS} photometry, compared to the previously published parameters. Notes: $^1$The closest mid-transit time to the midpoint of the \textit{CHEOPS} dataset; $^2$Adopted from literature; $^3$Derived based on $K = 131 \pm 29$ m~s$^{-1}$ \citep{Zhou1}. $^4$Derived based on $K < 352$ m~s$^{-1}$ \citep{Siverd1}. $^5$Derived based on $K < 400$ m~s$^{-1}$ \citep{Johnson1}. The best-fitting $P_\mathrm{orb}$ values are preliminary, see Table \ref{keltslineareph} for improved values.}
\label{photometryres}
\begin{tabular}{llll}
\hline
\hline
Parameter [unit] & Prior & This work & \citet{Zhou1}\\
\hline
\hline
\multicolumn{4}{c}{KELT-17b}\\
$T_\mathrm{c}$ [$\mathrm{BJD}_\mathrm{TDB}$] & $N$(2~459~215.9375, 0.0004)$^1$ & $2~459~215.937950 \pm 0.000086$ & $2~457~226.14219 \pm 0.00033$\\
$P_\mathrm{orb}$ [d] & $N$(3.0801718, 0.0000053)$^2$ & $3.0801724 \pm 0.0000047$ & $3.0801718 \pm 0.0000053$\\
$D$ & $N$(0.0085, 0.0001) & $0.008482 \pm 0.000049$ & $0.00907 \pm 0.00017$\\
$W$ (in phase units) & $N$(0.047, 0.001) & $0.04691 \pm 0.00013$ & $0.04701 \pm 0.00045$\\
$b$ (in units of stellar radius) & $N$(0.60, 0.03) & $0.587 \pm 0.011$ & $0.570 \pm 0.035$\\
$h_1$ & $N$(0.79, 0.10) & $0.7657 \pm 0.0094$ & --\\
$h_2$ (fixed) & 0.557 & 0.557 & --\\
\rprs & -- & $0.0921 \pm 0.0011$ & $0.09526 \pm 0.00088$\\  
$a/R_\mathrm{s}$ & -- & $6.246 \pm 0.077$ & $6.38 \pm 0.18$\\
$R_\mathrm{p}$ [$\mathrm{R}_\mathrm{Jup}$] & -- & $1.507 \pm 0.055$ & $1.525 \pm 0.065$\\
$M_\mathrm{p}$ [$\mathrm{M}_\mathrm{Jup}$] & -- & $1.31 \pm 0.29^3$ & $1.31 \pm 0.29$\\
$\log g_\mathrm{p}$ & -- & $3.154 \pm 0.099$ & $3.14 \pm 0.11$\\
$\rho_\mathrm{p}$ [$\mathrm{g.cm}^{-3}$] & -- & $0.47 \pm 0.11$ & $0.46 \pm 0.12$\\ 
\hline
\hline
Parameter [unit] & Prior & This work & \citet{Siverd1}\\
\hline
\hline
\multicolumn{4}{c}{KELT-19Ab}\\
$T_\mathrm{c}$ [$\mathrm{BJD}_\mathrm{TDB}$] & $N$(2~459~218.1780, 0.0005)$^1$ & $2~459~218.17799 \pm 0.00013$ & $2~457~281.24953 \pm 0.00036$\\
$P_\mathrm{orb}$ [d] & $N$(4.6117093, 0.0000088)$^2$ & $4.6117105 \pm 0.0000077$ & $4.6117093 \pm 0.0000088$\\
$D$ & $N$(0.0100, 0.0002) & $0.009702 \pm 0.000061$ & $0.01148 \pm 0.00020$\\
$W$ (in phase units) & $N$(0.038, 0.001) & $0.03791 \pm 0.00010$ & $0.03970 \pm 0.00032$\\
$b$ (in units of stellar radius) & $N$(0.55, 0.03) & $0.499 \pm 0.018$ & $0.601 \pm 0.030$\\
$h_1$ & $N$(0.79, 0.10) & $0.8064 \pm 0.0087$ & --\\
$h_2$ (fixed) & 0.542 & 0.542 & --\\
\rprs & -- & $0.0985 \pm 0.0010$ & $0.10713 \pm 0.00092$\\
$a/R_\mathrm{s}$ & -- & $8.213 \pm 0.088$ & $7.50 \pm 0.20$\\
$R_\mathrm{p}$ [$\mathrm{R}_\mathrm{Jup}$] & -- & $1.794 \pm 0.097$ & $1.91 \pm 0.11$\\
$M_\mathrm{p}$ [$\mathrm{M}_\mathrm{Jup}$] & -- & $<4.10^4$ & $<4.07$\\
$\log g_\mathrm{p}$ & -- & $<3.61$ & $<3.44$\\
$\rho_\mathrm{p}$ [$\mathrm{g.cm}^{-3}$] & -- & $<1.30$ & $<0.744$\\    
\hline
\hline
Parameter [unit] & Prior & This work & \citet{Johnson1}\\
\hline
\hline
\multicolumn{4}{c}{KELT-21b}\\
$T_\mathrm{c}$ [$\mathrm{BJD}_\mathrm{TDB}$] & $N$(2~459~055.3524, 0.0001)$^1$ & $2~459~055.352380 \pm 0.000086$ & $2~457~382.640727 \pm 0.00041$\\
$P_\mathrm{orb}$ [d] & $N$(3.6127647, 0.0000033)$^2$ & $3.6127640 \pm 0.0000031$ & $3.6127647 \pm 0.0000033$\\
$D$ & $N$(0.0100, 0.0001) & $0.009757 \pm 0.000054$ & $0.00990 \pm 0.00014$\\
$W$ (in phase units) & $N$(0.047, 0.001) & $0.04722 \pm 0.00013$ & $0.04734 \pm 0.00025$\\
$b$ (in units of stellar radius) & $N$(0.40, 0.01) & $0.4044 \pm 0.0095$ & $0.423 \pm 0.039$\\
$h_1$ & $N$(0.80, 0.10) & $0.7627 \pm 0.0089$ & --\\
$h_2$ (fixed) & 0.569 & 0.569 & --\\
\rprs & -- & $0.0987 \pm 0.0011$ & $0.09952 \pm 0.00073$\\
$a/R_\mathrm{s}$ & -- & $6.885 \pm 0.081$ & $6.86 \pm 0.13$\\
$R_\mathrm{p}$ [$\mathrm{R}_\mathrm{Jup}$] & -- & $1.610 \pm 0.034$ & $1.586 \pm 0.040$\\
$M_\mathrm{p}$ [$\mathrm{M}_\mathrm{Jup}$] & -- & $<3.70^5$ & $<3.91$\\
$\log g_\mathrm{p}$ & -- & $<3.32$ & $<3.59$\\
$\rho_\mathrm{p}$ [$\mathrm{g.cm}^{-3}$] & -- & $<1.18$ & $<1.24$\\
\hline
\hline
\end{tabular}
\end{table*}

To derive the system parameters, we used the dedicated \textit{CHEOPS} transit analysis software called {\tt{pycheops}}\footnote{See \url{https://github.com/pmaxted/pycheops}.} \citep{Maxted3}. This {\tt{Python}}\footnote{See \url{https://www.python.org/}.} package includes downloading, visualizing, and decorrelating \textit{CHEOPS} data, fitting transits and eclipses of exoplanets, and calculating light curve noise. We first cleaned the light curves from outlier data-points using the {\tt{pycheops}} built-in function {\tt{clip\_outliers}}, which removes outliers from a dataset by calculating the mean absolute deviation ($MAD$) from the light curve following median smoothing, and rejects data greater than the smoothed dataset plus the $MAD$ multiplied by a clipping factor. The clipping factor equal to five was reasonable in our cases, which we checked visually. The next step was the roll angle effect subtraction. In order to keep the cold plate radiators facing away from the Earth, the spacecraft rolls during its orbit. This causes that the field of view rotates around the pointing direction. The target star remains stationary within typically a pixel, but the rotation of the field of view produces a variation of its flux from the nearby sources in phase with the roll angle of the spacecraft \citep{Bonfanti1}. At exoplanet transits this is usually a small fraction of the total flux, in a form of short-term, non-astrophysical flux trends -- waves or bumps. This rolling effect is not corrected by the DRP, but it is possible to perform this correction using the {\tt{pycheops}} function called {\tt{decorr}}. By this "derolling" procedure the $RMS$ of the light-curve residuals decreased by about 30 - 50 ppm. Subsequently, the residuals were visually checked against the roll angle to ensure that the removal of rolling systematics has been done properly. Using this function it is possible to model first, second or third order trends in the flux over time, x or y centroid, roll angle, background, or contamination. After the decorrelation process the data are ready for model fitting.

We first fitted the transit light curves individually. The {\tt{pycheops}} package uses the {\tt{qpower2}} transit model with the power-2 limb darkening law \citep{Maxted2, Maxted1}. Transit models are constructed using the following transit parameters: the orbital period $P_\mathrm{orb}$, the mid-transit time $T_\mathrm{c}$, the transit depth $D$, which is defined as $D = (R_\mathrm{p}/R_\mathrm{s})^2$, where $R_\mathrm{p}/R_\mathrm{s}$ is the planet-to-star radius ratio, the transit width $W$ (in phase units), the impact parameter $b$ (in units of stellar radius), which is defined as $b = a \cos i/R_\mathrm{s}$, where $a$ is the semi-major axis of the planet's orbit and $i$ is the orbit inclination angle with respect to the plane of the sky, the flux scaling factor $c$, the limb-darkening coefficients $h_1$ and $h_2$, and the orbital eccentricity and longitude of periastron components $f_\mathrm{c} = e \cos \omega/\sqrt e$ and $f_\mathrm{s} = e \sin \omega/\sqrt e$, where $e$ is the eccentricity and $\omega$ is the longitude of periastron. Several decorrelation parameters are also used. During the individual transit analysis we fixed the orbital period $P_\mathrm{orb}$ using the already published literature values \citep{Zhou1, Siverd1, Johnson1}. We also did not adjust the limb-darkening coefficient $h_2$, which was interpolated from the stellar parameters ($T_\mathrm{eff}$, $\log g$, and Fe/H), tabulated in the {\tt{SWEET-Cat}} database \citep{Santos1} using the {\tt{ATLAS}} model, see e.g., \citet{Claret1}. We note that prior this treatment we ran several test modelings with $h_2$ allowed to float, however, we always got unphysical fitted coefficient far from the interpolated value and the corresponding fit was inappropriate. Very probably this is due to the high $T_\mathrm{eff}$ of the planet hosts. Finally, we decided to keep fixed this coefficient, which reflectes the average temperatures of the stars. The average temperatures are closer to the lowest temperature, because the polar star regions, where is the highest effective temperature, have the smallest area. We assumed circular orbit for KELT-17b, KELT-19Ab, and KELT-21b, thus the $f_\mathrm{c}$ and $f_\mathrm{s}$ parameters were set to zero. Other parameters were freely adjusted.    

The final parameters were derived using a Markov chain Monte Carlo (MCMC) methodology. The software {\tt{pycheops}} does this by utilizing the affine invariant sampler {\tt{Python}} package {\tt{emcee}}\footnote{See \url{https://emcee.readthedocs.io/en/stable/}.} \citep{Foreman1} to sample the posterior probability distribution of fitting the constructed transit model to the data. The best fit values from the {\tt{qpower2}} analysis are used as priors for the {\tt{emcee}} sampler function. The sampler has built-in functionality to fit and remove correlated stellar noise using a Gaussian process regression method from the {\tt{celerite2}}\footnote{See \url{https://celerite.readthedocs.io/en/stable/}.} package \citep{Kallinger1, Foreman2, Barros1}. The regression is done by using a {\tt{SHOTerm}} plus {\tt{JitterTerm}} kernel, with a fixed quality factor $Q = 1/\sqrt{2}$, implemented in the {\tt{celerite2}} package. It uses $\log \sigma$ (free), $\log Q$ (fixed), $\log \omega_0$ (free), and $\log S_0$ (free) hyperparameters with bounds on the values of these parameters to be inputted by the user. We first fixed the transit shape, i.e., the parameters $D$, $W$, and $b$, and the mid-transit time $T_\mathrm{c}$ from the initial {\tt{qpower2}} fit and set free the three hyperparameters for a preliminary MCMC analysis. The posteriors of the hyperparameters obtained from this analysis were used to define the priors for the next MCMC analysis as twice the uncertainty computed from the posterior distribution. Finally, we ran the MCMC analysis again with free transit model parameters and free hyperparameters.

\subsection{Joint transit analysis}
\label{jointanalysis}

To combine the best-fitting results obtained from the individual \textit{CHEOPS} light curves and to get final parameters of the exoplanet systems, we performed a joint analysis of the dataset per object using the \texttt{pycheops} package. Because the observations were obtained at multiple epochs, in this case we fitted not only the transit shape, i.e., the parameters $D$, $W$, and $b$, and the mid transit time $T_\mathrm{c}$, but also the orbital period $P_\mathrm{orb}$ of the planet. As during the individual analysis, we fitted only the limb-darkening coefficient $h_1$ and fixed $h_2$ as interpolated from the stellar parameters, and we also fixed the $f_\mathrm{c}$ and $f_\mathrm{s}$ parameters to zero. We also used the decorrelation parameters of each single visit, and the Gaussian process regression method from the \texttt{celerite2} package, with the common hyperparameters of $\log \omega_0$ and $\log S_0$. The priors on the hyperparameters were determined as the average (with error propagation) of the single-visit hyperparameters. We note that in the joint transit analysis mode, the roll angle model is not part of the detrending model as in the individual transit analysis mode. The detrending parameters of the roll angle (and its harmonics) are treated as nuisance parameter \citep{Luger1} and they are marginalized away as a \texttt{celerite2 CosineTerm} kernel added to the covariance matrix. This method implicitly assumes that the roll angle is a linear function of time for each visit. The results obtained from this joint transit analysis are presented in Sect. \ref{parameters}.

As the next step we took the best fitting parameters of $T_\mathrm{c}$ and $P_\mathrm{orb}$ from the joint analysis and used them as fixed parameters for calculating transit time variations based on the joint model. In this case the \texttt{pycheops} software fits the transit shape as previously during the joint transit analysis, but in addition, it also fits the deviation $\Delta T_\mathrm{0,n}$ for the $n$-th visit from the calculated individual mid-transit time. The observed mid-transit time $T_\mathrm{0,n}$ is defined based on the linear ephemeris as:

\begin{equation}
\label{ttvpycheops}
T_\mathrm{0,n} = T_\mathrm{c} + P_\mathrm{orb} \times E + \Delta T_\mathrm{0,n},
\end{equation} 
     
\noindent where $E$ is the epoch of observation, i.e., the number of the orbital cycle. The fitted $\Delta T_\mathrm{0,n}$ value corresponds to the observed-minus calculated (O-C) value of mid-transit time, which is a very effective tool to reveal transit time variations of planets through the O-C diagram. The \textit{CHEOPS} space telescope can be used for this purpose as it was recently discussed by \citet{Borsato1}. The transit timing analysis of the data is detailed in Sect. \ref{ttv}.

\subsection{Overview and discussion of the refined system parameters}      
\label{parameters}

We summarize the fitted and derived parameters of the planetary systems in Table \ref{photometryres}. We also present the previously published parameters for easy comparison. The phase-folded transit light curves of the exoplanets KELT-17b, KELT-19Ab, and KELT-21b, overplotted with the best-fitting \texttt{pycheops} models are presented in Figs. \ref{kelt1719fittedlc} and \ref{kelt21fittedlc}. 

\subsubsection{KELT-17b}

Based on the \textit{CHEOPS} observations, the planet KELT-17b is a close-in hot Jupiter with an orbital period of $P_\mathrm{orb} = 3.0801724 \pm 0.0000047$ d. The orbital period was improved further in Sect. \ref{ttv}. It is a massive, inflated planet, its mass is $M_\mathrm{p} = 1.31 \pm 0.29$ $\mathrm{M}_\mathrm{Jup}$ and its radius is $R_\mathrm{p} = 1.507 \pm 0.055$ $\mathrm{R}_\mathrm{Jup}$, which gives the planet density of $\rho_\mathrm{p} = 0.47 \pm 0.11$ $\mathrm{g.cm}^{-3}$. This value is only about 35\% of the Jupiter's density. Based on the \textit{CHEOPS} measurements the planet body seems to be smaller in comparison with the value presented by the discoverers. The fitted transit depth is $D = 0.008482 \pm 0.000049$. \citet{Zhou1} obtained the transit depth of $D = 0.00907 \pm 0.00017$, which is about $3.4\sigma$ difference. The planet-to-star radius ratio parameter was derived from the transit depth, giving the value of \rprs = $0.0921 \pm 0.0011$. This is almost $3\sigma$ difference in comparison with the value of \rprs = $0.09526 \pm 0.00088$, presented by the discoverers. This is an interesting result, suggesting that the planet is not so inflated as found before. On the other hand, this also could be due either to the difference in spectral response of the applied detectors, or the reason could be a parameter degeneracy between $D$ and $b$. The disadvantage of \textit{CHEOPS} observations from this viewpoint is the lack of multicolor data. Other fitted and derived parameters are in a $3\sigma$ agreement with the discovery paper, but we improved several parameter values in comparison with \citet{Zhou1}, for example in the case of $a/R_\mathrm{s}$ with a factor of 2.3, or in the case of $T_\mathrm{c}$ with a factor of 3.8.

\subsubsection{KELT-19Ab}
\label{k19Abcheops}

KELT-19Ab is a close-in giant hot-Jupiter-type planet with an orbital period of $P_\mathrm{orb} = 4.6117105 \pm 0.0000077$ d (see Table \ref{keltslineareph} for the improved value). This parameter value was derived based on the four \textit{CHEOPS} observations and it is in a $3\sigma$ agreement with the orbital period found by the discoverers. Other fitted parameters are, however, significantly different in comparison with the parameter values presented by \citet{Siverd1}. This indicates that the parameter degeneracy between $D$ and $b$ as the reason for this discrepancy is less probable. The transit is shallower, we obtained a transit depth of $D = 0.009702 \pm 0.000061$, which is almost $9\sigma$ difference in comparison with the previously obtained value of $D = 0.01148 \pm 0.00020$. Consequently, the planet body is also smaller, the derived parameter \rprs is $0.0985 \pm 0.0010$, where is about $8.6\sigma$ difference in comparison with the value presented by the discoverers. Note that we could not refine only this parameter using \textit{CHEOPS} observations. The obtained impact parameter is also very different, i.e., we obtained $b = 0.499 \pm 0.018$, while \citet{Siverd1} derived $b = 0.601 \pm 0.030$, which differs by about $3.4\sigma$ from our value. The scaled semi-major axis $a/R_\mathrm{s}$ also seems to be significantly larger by about $3.5\sigma$. The telescope rotation cannot cause such a discrepancy, mainly because the rolling effect is too small, moreover, beacause the observed flux was decorrelated against the roll angle. The mass of the planet is not constrained well due to the scatter in the discovery radial velocity measurements, as well as the parameters derived from the planet mass are only upper limited (see Table \ref{photometryres}). The last parameter, i.e., the transit width $W$ (the transit duration) is discussed in Sect. \ref{tdv}. 

\citet{Yang1} recently reported on results of a follow-up photometry observation of three exoplanets, including KELT-19Ab, using precise \textit{TESS} data (see Table \ref{TESSobslog}). \citet{Yang1} analyzed 2-minute cadence data of KELT-19Ab and they corrected the contamination coming from the field stars using the Gaia database \citep{Gaiadr2}. The authors derived the following system parameters: \rprs = $0.09955 \pm 0.00074$, $i = 88.9 \pm 0.7$ deg, and $a/R_\mathrm{s} = 9.10 \pm 0.19$. We can see that the planet-to-star radius ratio value derived from the \textit{TESS} data is comparable with the \rprs value obtained based on the \textit{CHEOPS} measurements (the difference is about $1\sigma$). On the other hand, the orbit inclination angle value\footnote{Converting the value of $b = 0.499 \pm 0.018$, obtained based on the \textit{CHEOPS} data, to orbit inclination angle gives $i = 86.17 \pm 0.14$ deg.} $i$ and the scaled semi-major axis value $a/R_\mathrm{s}$ is significantly different from the \textit{CHEOPS}-based values. 

Furthermore, we reanalyzed the mentioned \textit{TESS} 2-minute cadence dataset of KELT-19Ab using the \texttt{RMF} code, as it is described in Sect. \ref{tdv}. We got \rprs = $0.09550 \pm 0.00030$, $i = 87.65 \pm 0.36$ deg, and $a/R_\mathrm{s} = 8.70 \pm 0.17$, see Table \ref{rmfres2} for further parameter values. These results also indicate that the size of KELT-19Ab is smaller than originally derived by the discoverers and that its orbital parameters are different, too.

\subsubsection{KELT-21b}

KELT-21b is a hot Jupiter-type planet with an orbital period of $P_\mathrm{orb} = 3.6127640 \pm 0.0000031$ d, improved further in Sect. \ref{ttv}, orbiting the host KELT-21, which is the most rapidly rotating star to host a transiting planet. Based on the newly obtained spectra (see Sect. \ref{spectra}) we derived the $v \sin I_* = 141.9 \pm 2.4$ km~s$^{-1}$, but \citet{Johnson1} presented a more precise value of $v \sin I_* = 146.03 \pm 0.48$ km~s$^{-1}$. The discoverers announced a pair of faint stars at a projected separation of $1.2''$ from the host, which could be a pair of stars bound with KELT-21 in a triple system. If confirmed in the future, the KELT-21 system would be very unique, with two M dwarfs and a fast rotating A-type planet host. 

Based on the \textit{CHEOPS} observations we significantly improved the system parameters (except for the planet-to-star-radius ratio \rprs parameter) in comparison with \citet{Johnson1}, e.g., the impact parameter $b$ with a factor of 4.1, or the mid-transit time $T_\mathrm{c}$ with a factor of 4.7. Every parameter derived based on the \textit{CHEOPS} observations is in a $3\sigma$ agreement with the corresponding parameter value presented by the discoverers. Based on the improved parameters, KELT-21b is a massive Jupiter-size planet with $R_\mathrm{p} = 1.610 \pm 0.034$ $\mathrm{R}_\mathrm{Jup}$, transiting the host in a distance of $a = 6.885 \pm 0.081$ $R_\mathrm{s}$ with the impact parameter of $b = 0.4044 \pm 0.0095$, causing the transit depth of $D = 0.009757 \pm 0.000054$. Due to the large scatter in the discovery radial velocity measurements, caused by rapid rotation of KELT-21, we can estimate only the upper limit on the planet's mass as $M_\mathrm{p} < 3.70$ $\mathrm{M}_\mathrm{Jup}$. Similarly, the derived parameters of $\log g_\mathrm{p} < 3.32$ cgs and $\rho_\mathrm{p} < 1.18$ $\mathrm{g.cm}^{-3}$ are also upper limited only, similarly as in the case of KELT-19Ab (see Table \ref{photometryres}).                         

\section{The \textit{CHEOPS} transit light curves from the viewpoint of spin-orbit misalignment}  
\label{spinorb}

\begin{table*}
\centering
\caption{An overview of the \rmf best-fitting parameters of the exoplanets KELT-17b, KELT-19Ab, and KELT-21b obtained from the \textit{CHEOPS} photometry and using the gravity-darkening approach of \citet{Espinosa1}. Notes: $^1$The closest mid-transit time to the midpoint of the \textit{CHEOPS} dataset. $^2$Based on \texttt{pycheops} results. $^3$Assuming circular orbit. $^4$Based on the literature values \citep{Zhou1, Siverd1, Johnson1}. $^5$Minimum value, assuming $I_* = 90$ deg. $^6$Unknown parameter, assuming $I_* = 90$ deg.}
\label{rmfres}
\begin{tabular}{llll}
\hline
\hline
Parameter [unit] & KELT-17b & KELT-19Ab & KELT-21b\\
\hline
\hline
$T_\mathrm{c}$ [$\mathrm{BJD}_\mathrm{TDB}$]$^1$ & $2~459~215.937912 \pm 0.000073$  & $2~459~218.17800 \pm 0.00012$ & $2~459~055.35251 \pm 0.00016$\\
$P_\mathrm{orb}$ [d] (fixed)$^2$ & $3.0801724$ & $4.6117105$ & $3.6127640$\\
$i$ [deg] & $84.780 \pm 0.071$ & $88.66 \pm 0.33$ & $87.19 \pm 0.13$\\
$R_\mathrm{s}/a$ & $0.15952 \pm 0.00068$ & $0.1100 \pm 0.0014$ & $0.14357 \pm 0.00090$\\
\rprs & $0.09186 \pm 0.00016$ & $0.09640 \pm 0.00021$ & $0.09551 \pm 0.00016$\\
$e$ (fixed)$^3$ & $0.0$ & $0.0$ & $0.0$\\
$\omega$ [deg] (fixed)$^3$ & $90.0$ & $90.0$ & $90.0$\\
$\lambda$ [deg] (0 deg if aligned; fixed)$^4$ & $244.0$ & $180.3$ & $354.4$\\
$\mathrm{d}i/\mathrm{d}t$ [deg d$^{-1}$] (fixed) & $0.0$ & $0.0$ & $0.0$\\
$\Omega/\Omega_\mathrm{crit}$ (fixed)$^5$ & $0.113$ & $0.230$ & $0.394$\\
$I_*$ [deg] (fixed)$^6$ & $90.0$ & $90.0$ & $90.0$\\
$l_\mathrm{norm}$ & $1.0000750 \pm 0.0000010$ & $1.0000740 \pm 0.0000010$  & $1.0000760 \pm 0.0000010$\\
$l_3$ (fixed) & $0.0$ & $0.0$ & $0.0$\\
\hline
\hline
\end{tabular}
\end{table*}

\begin{figure*}
\centering
\centerline{
\includegraphics[width=\columnwidth]{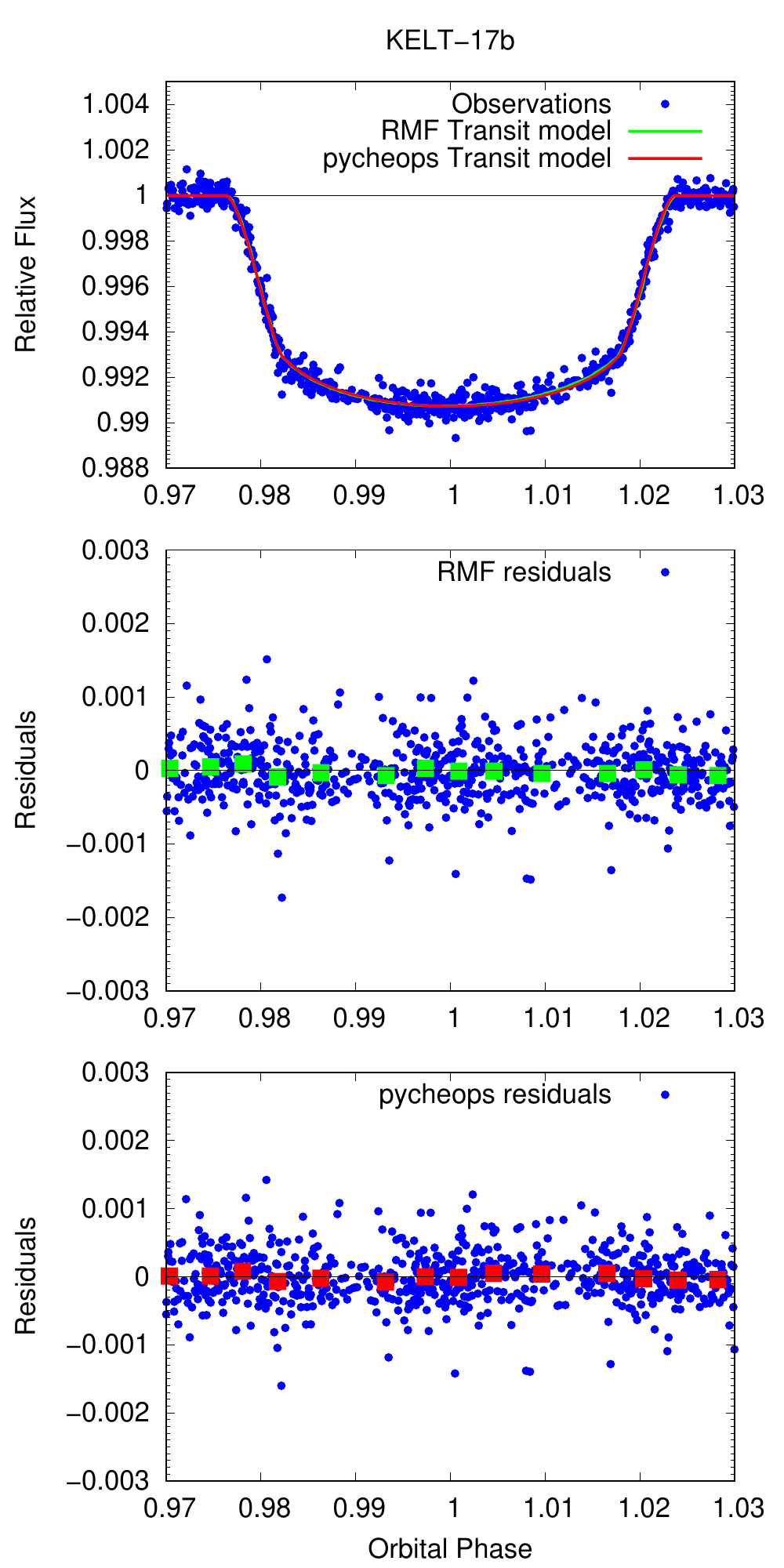}
\includegraphics[width=\columnwidth]{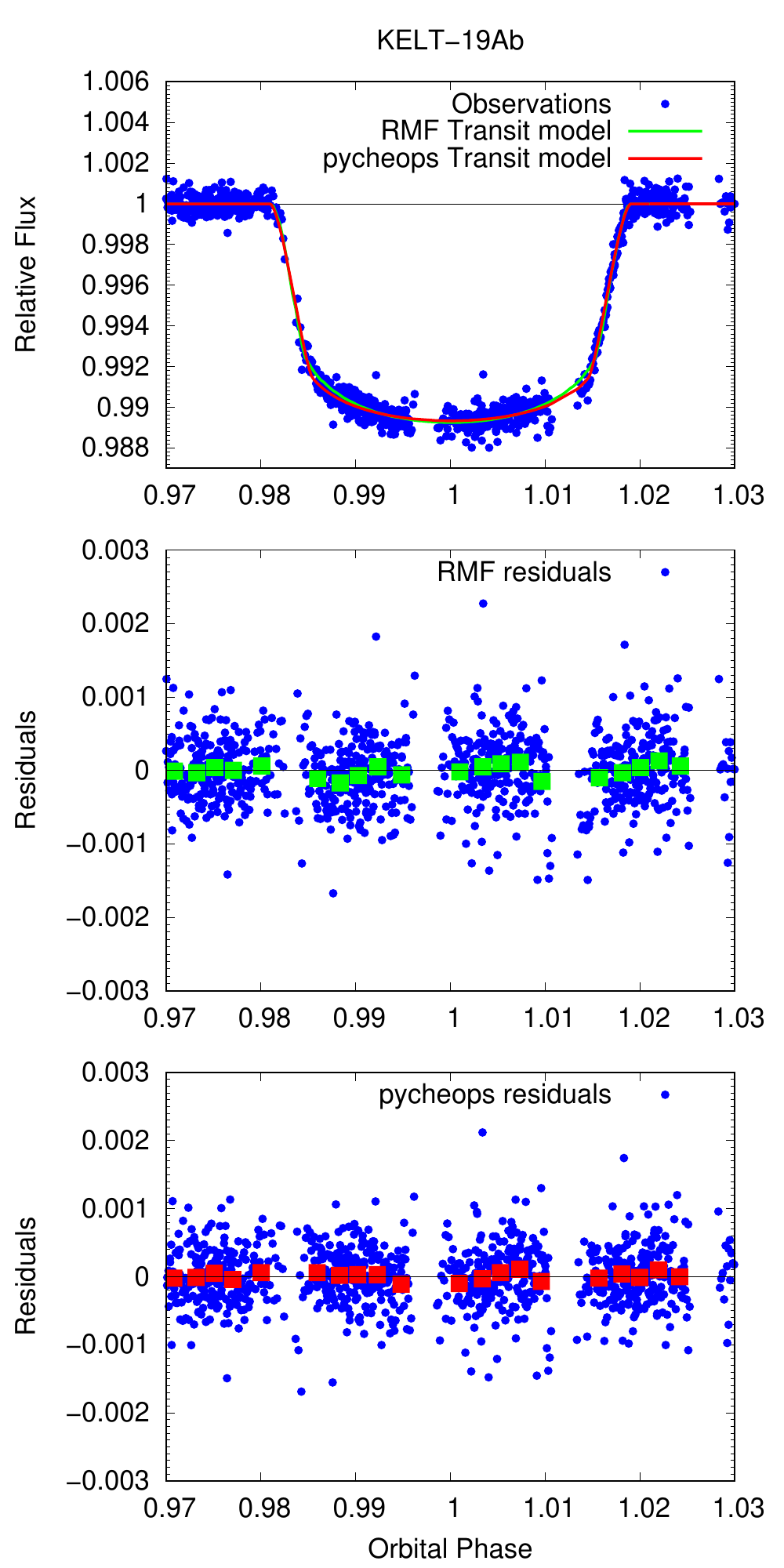}}
\caption{Phase-folded transit light curves of KELT-17b (top left-hand panel) and KELT-19Ab (top right-hand panel), overplotted with the best-fitting \rmf models. The corresponding residuals are also shown (middle panels). The graphs were cropped to focus on transit events, where the asymmetry is expected. The best-fitting \texttt{pycheops} transit models (top panels) and residuals (bottom panels) are copied here for comparison purposes. The residuals were binned to highlight the possible wave shape (1 bin-point represents 50 data-points).}
\label{kelt1719fittedlcrmf} 
\end{figure*} 

\begin{figure}
\includegraphics[width=\columnwidth]{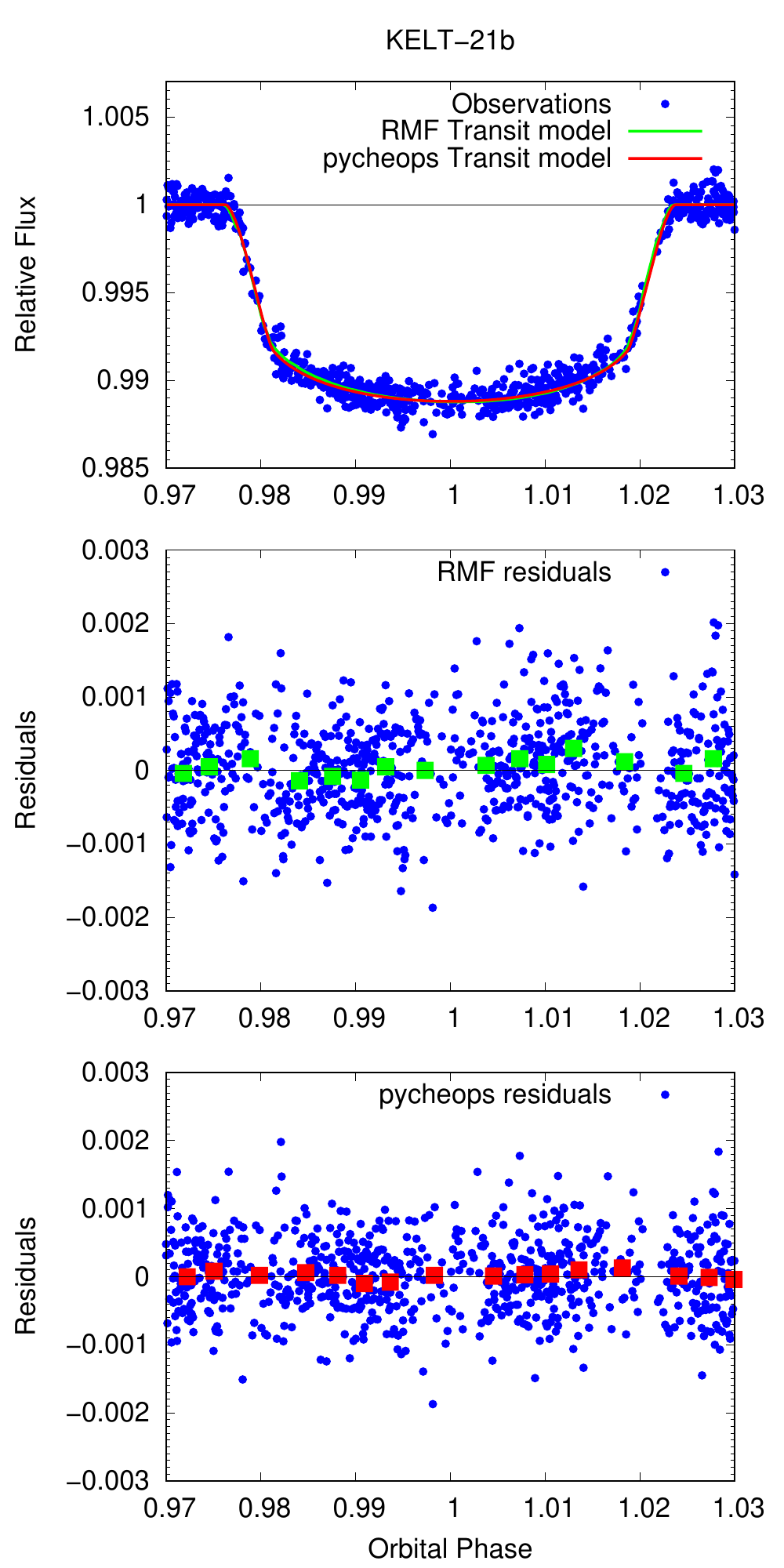}
\caption{As in Fig. \ref{kelt1719fittedlcrmf}, but for KELT-21b.}
\label{kelt21fittedlcrmf} 
\end{figure}

Rossiter-McLaughlin effect \citep{Rossiter1, McLaughlin1} observations revealed several spin-orbit misaligned planets, see e.g., \citet{Narita1}, \citet{Hebrard1}, or \citet{Johnson2}. This technique can determine only the sky-projected spin-orbit misalignment angle $\lambda$, moreover due to the rapid rotation of early-type stars the radial velocity measurements are challenging in these cases. Therefore, at fast rotators the transit photometry method \citep{Barnes1} is used instead of radial velocities, if the data are precise enough, see e.g., \citet{Szabo1}. The advantage of this method is that it is also possible to derive the stellar inclination $I_*$, and thus the true misalignment \citep{Barnes2}. Several \textit{CHEOPS} observations were also used for this purpose, for example in \citet{Lendl1}. Since the effect of rapid rotation in transit light curves are maximized at short wavelengths \citep{Barnes1}, the advantage of the \textit{CHEOPS} observatory compared, e.g., with the \textit{TESS} telescope is the bluer spectral window of the \textit{CHEOPS} instrument\footnote{See \url{https://www.cosmos.esa.int/web/cheops/performances-bandpass}.}. We also aimed at searching for photometric indicators of spin-orbit misalignment in the precise \textit{CHEOPS} transit light curves of KELT-17b, KELT-19Ab, and KELT-21b, therefore we tested the obtained \textit{CHEOPS} data from the viewpoint of transit asymmetry. For this purpose we used the same DRP processed "OPTIMAL" light curves as in Sect. \ref{sysparamsfromcheops}. The data were detrended as it is described in Sect. \ref{indanalysis}, i.e., using the \texttt{pycheops} function \texttt{decorr}, where mainly the rolling effect of the telescope is removed, but in this case we did not use the Gaussian process regression method to avoid overcompensation of the light curves \citep{Borsato1}. We wanted to preserve the possible transit asymmetry with this data treatment. 

The detrended \textit{CHEOPS} transit data were analyzed using the \rmf (Roche ModiFied) code. The software was prepared based on the {\tt{ROCHE}} code, which is devoted to the modeling of multi-data set observations of close eclipsing binary stars, such as radial velocities and multi-color light curves \citep{Pribulla2}. The \rmf code was already used with success, e.g., in \citet{Szabo3}, where the spin-orbit misaligned Kepler-13A system were re-analyzed using \textit{Kepler} and \textit{TESS} data. The software can simultaneously model multi-color light curves, radial velocities, and broadening functions, or least-squares deconvolved line profiles of binary stars and transiting exoplanets. Its modification to be used with the transiting exoplanets uses the Roche surface geometry with the planet gravity neglected for the host star (rotationally deformed shape) and spherical shape for the planet. The model can handle eccentric orbits, misaligned rotational axes of the components, stellar oblateness, gravity darkening due to rapid rotation using the analytical approach of \citet{Espinosa1}\footnote{This model assumes that the latitudinal variation of $T_\mathrm{eff}$ only depends on a single parameter, namely the ratio of the equatorial velocity to the Keplerian velocity (the gravity darkening exponent is removed).}, Doppler beaming effect, advanced limb-darkening description, and third light. The synthesis of the broadening functions assumes solid-body rotation. The synthesis of the observables is performed in the plane of the sky using pixel elements. The effectiveness of the integration is increased by the adaptive phase step being more fine during the eclipses/transits.

The software uses the following parameters: the mid-transit time $T_\mathrm{c}$, the orbital period $P_\mathrm{orb}$, the orbit inclination angle $i$ with respect to the plane of the sky, the ratio of the host radius to the semi-major axis $R_\mathrm{s}/a$, the eccentricity $e$, the longitude of the periastron passage $\omega$, the sky-projected spin-orbit misalignment angle $\lambda$, the orbit inclination angle change rate $\mathrm{d}i/\mathrm{d}t$, the inclination angle of the stellar rotation axis $I_*$, the planet-to-star radius ratio $R_\mathrm{p}/R_\mathrm{s}$, the third light $l_3$, defined as $l_3/(l_1+l_2)$, the light-curve normalization factor $l_\mathrm{norm}$, and the ratio of the stellar angular rotation velocity to the break-up velocity $\Omega/\Omega_\mathrm{crit}$, which defines the rotationally deformed stellar shape and the temperature distribution on the stellar surface, see Eq. 1 in \citet{Szabo3}. The stellar limb darkening is described by the four-parameter model of \citet{Claret1} with the critical foreshortening angle approach. The limb-darkening coefficients ($a_1$, $a_2$, $a_3$, $a_4$, and $\mu_\mathrm{crit}$)\footnote{Parameter $\mu = \cos \theta$, where $\theta$ is the so-called foreshortening angle, which is angle between the line of sight and a normal to the stellar surface. For $\mu < \mu_\mathrm{crit}$ the stellar flux is assumed to be zero, see \citet{Claret1}.} were calculated for the \textit{CHEOPS} passband using the same spherical {\tt{PHOENIX-COND}} models as in \citet{Claret1}. The applied coefficients were linearly interpolated from the calculated table for the local gravity and temperature for each surface pixel, based on the already published Fe/H parameter values, listed in Table \ref{spectrares}. This is important, because the local gravity and the effective temperature vary due to the stellar rotation. The local values of temperature were calculated using the approach of \citet{Espinosa1} from the local gravity. The values of the polar temperature and gravity were adjusted, so the mean value of the surface distribution was close to the already published stellar parameters of $T_\mathrm{eff}$ and $\log g$, presented in Table \ref{spectrares}. These coefficients were fixed during the fitting procedure. We also kept fixed the orbital period based on the \texttt{pycheops} results, and the $\lambda$ values as presented in the literature \citep{Zhou1, Siverd1, Johnson1}. In addition, the eccentricity $e$ was set to zero and the longitude of the periastron passage $\omega$ was fixed at $90$ deg, i.e., we assumed circular orbit of the exoplanets. We assumed no change in the orbit inclination angle with time, i.e., we set $\mathrm{d}i/\mathrm{d}t$ to zero. The parameter $\Omega/\Omega_\mathrm{crit}$ was calculated based on the stellar mass $M_*$ and radius $R_*$, presented in the literature, see Table \ref{spectrares}, and then it was fixed during the fitting procedure. We calculated the star's polar radius based on its mean radius using the Eq. 16 of \citet{Zahn1}. Since $I_*$ is unknown parameter, even in the case of KELT-17 is very uncertain, we assumed $I_* = 90$ deg, thus we could calculate only the minimum value of $\Omega/\Omega_\mathrm{crit}$. Finally, we also fix the $l_3$ parameter to zero, because the third light contamination was removed by the DRP. 

\begin{figure*}
\centering
\includegraphics[width=175mm]{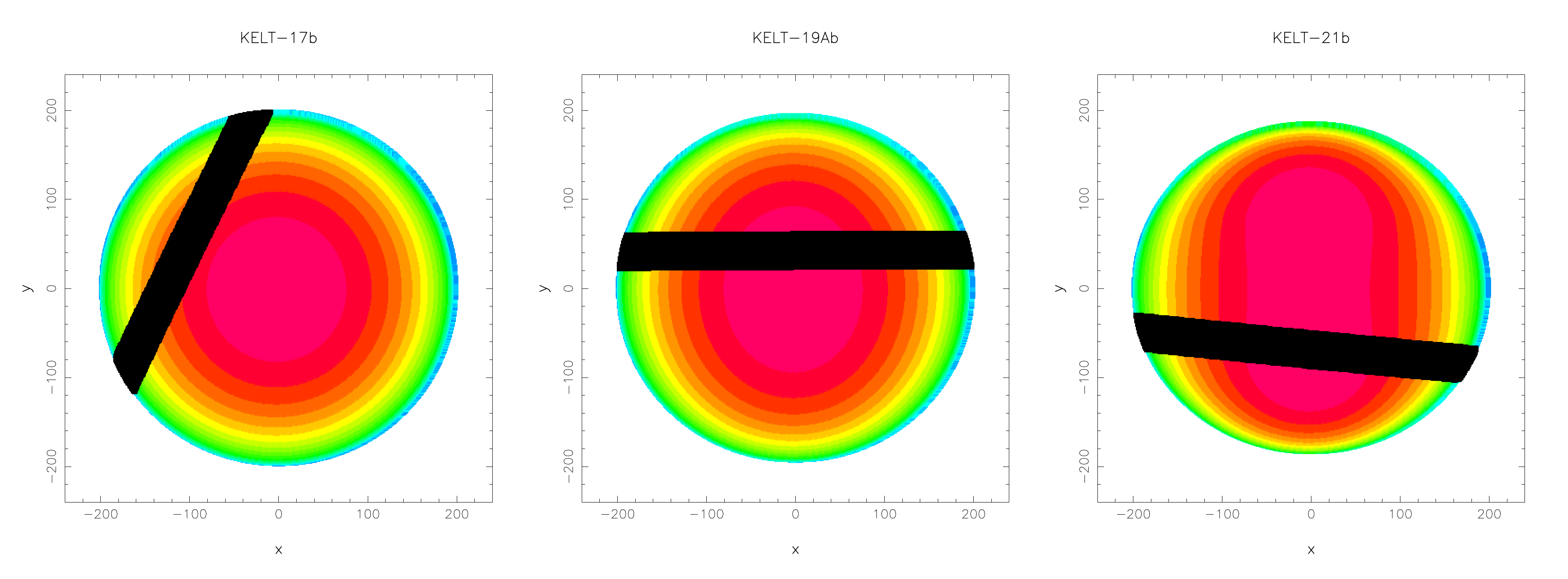}
\caption{2D illustration of the transit chord of KELT-17b, KELT-19Ab, and KELT-21b in front of the stellar surfaces. The intensity distribution was calculated for the effective wavelength of the \textit{CHEOPS} instrument. The green-yellow-red-magenta coloring scheme of the stellar surfaces reflects the increasing local flux as seen by the observer. The stellar inclination is always $I_* = 90$ deg.}
\label{systems2D}
\end{figure*}

We present the best-fitting \rmf parameters of KELT-17, KELT-19A, and KELT-21 systems in Table \ref{rmfres}. The uncertainties in the fitted parameters were derived based on the covariance matrix method. The majority of the parameter values corresponds to the \texttt{pycheops} parameter solutions within $3\sigma$. The phase-folded transit light curves of the exoplanets are depicted in Figs. \ref{kelt1719fittedlcrmf} and \ref{kelt21fittedlcrmf}. We can see that precision of the \textit{CHEOPS} observations is insufficient to conclude on spin-orbit misalignment of the planets. The point-to-point scatter is too big to detect the transit asymmetry in the data, which is about 50, 10, and 150 ppm for KELT-17b, KELT-19Ab, and KELT-21b, respectively. The \rmf and \texttt{pycheops} models are close to each other and both satisfy the observations (see top panels in Figs. \ref{kelt1719fittedlcrmf} and \ref{kelt21fittedlcrmf}). We did not detect the expected wave shape in the residuals, even though we binned the data points to highlight the possible asymmetry. This no detection is due to the characteristics of the systems. We at least confirmed that the gravity-darkening effect is very low in these cases. In the case of KELT-17b (see Fig. \ref{systems2D} left-hand panel) the spin-orbit misalignment angle is $\lambda = 244.0$ deg, which is in favor of the gravity-darkening effect, but on the other hand, the host star rotates relatively slowly with $v \sin I_* = 48.49 \pm 0.15$ km~s$^{-1}$. KELT-19A rotates faster with $v \sin I_* = 86.36 \pm 0.21$ km~s$^{-1}$, but the planet KELT-19Ab is in a retrograde orbit regime (see Fig. \ref{systems2D} middle panel), which causes very low gravity-darkening effect. The third system in our sample, KELT-21, was the most promising due to the very fast rotating host star with $v \sin I_* = 141.9 \pm 2.4$ km~s$^{-1}$, however the low gravity-darkening effect is caused by the almost aligned orbit regime of KELT-21b (see Fig. \ref{systems2D} right-hand panel). We can conclude that more precise observations are needed to detect these fine effects in the future. Furthermore, shorter wavelegths observations than the \textit{CHEOPS} spectral window can make the detection easier. Finally, we tested the effect of a change in the inclination angle of the stellar rotation axis $I_*$ on the quality of the fit. We set free the $I_*$ parameter and measured the quality of the fit when $I_*$ is changing. The $\chi^2$ parameter was used as a goodness-of-fit indicator. In the cases of the relatively slowly rotating systems of KELT-17 and KELT-19A it does not affect the quality of the fit, thus we could not draw any conclusions for these systems. At the very fast rotating KELT-21 system there is an indication that the inclination angle of the stellar rotation axis is $I_* \approx 60$ deg. We registered the lowest $\chi^2$ at this value, and the quality of the fit decreased both below and above of this inclination angle value.       

\section{Search for transit duration variations and transit time variations in the systems}
\label{tdvttv}         

Transit duration variations (TDVs) are possible in planetary systems with rapidly rotating host stars. In the case of the ''prototype'' Kepler-13A system the identified long-term TDV is caused by precession of the orbital plane of the exoplanet Kepler-13Ab. The orbital precession is induced by oblateness of the host star. \citet{Szabo2} found that the duration of Kepler-13Ab transits is gradually increasing with a rate of $(1.14 \pm 0.30) \times 10^{-6}$ d~cycle$^{-1}$. Moreover, the authors suggested that the reason for this variation is the expected change of the impact parameter with a rate of $\mathrm{d}b/\mathrm{d}t = -0.016 \pm 0.004$ yr$^{-1}$. Later, the orbital precession was confirmed by \citet{Masuda1}. \citet{Szabo3} revisited the impact parameter change rate using available \textit{Kepler} and \textit{TESS} data, and found a value of $\mathrm{d}b/\mathrm{d}t = -0.011$ yr$^{-1}$. In this part of our work, we also investigated the presence of long-term TDV in KELT-17, KELT-19A, and KELT-21 systems, searching for possible orbital precession. Based on Eq. 12 in \citet{Szabo2} the orbital precession probability is higher in the case of KELT-17b, where $\lambda = 244.0$ deg, and lower in the cases of KELT-19A and KELT-21, where nearly retrograde and aligned orbit was identified, respectively \citep{Zhou1, Siverd1, Johnson1}.        

Transit time variations (TTVs) were identified only in the case of a few hot Jupiters. The main reason for these variations is the suspected outer companions, i.e., planets and brown dwarfs, see for example \citet{Dawson1}, \citet{Nascimbeni1}, \citet{Maciejewski1}, \citet{Knutson1}, \citet{Hartman1}, \citet{Neveu-VanMalle1}, or \citet{Gajdos1}. Long-term TTVs were considered due to the tidal decay, for example by \citet{Hellier1}, \citet{Oberst1}, \citet{Gillon1, Gillon2}, or by \citet{Hebb1}, however up to now WASP-12b is the only hot Jupiter to have a decaying orbit confirmed \citep{Turner1}. These examples are, however, relatively rare to the number of known hot Jupiters. Some known examples of TTVs has recently been debated, see for example \citet{Seeliger1}, \citet{Wang1}, or \citet{Turner2}. \citet{Szabo3} also searched for possible TTVs in the rapidly rotating Kepler-13A system. In this particular case the absence of any TTVs is very strongly constrained by the two sources of data, i.e., \textit{Kepler} and \textit{TESS}. Rapid rotation as the primary reason for the TTV signal was not confirmed within hot Jupiter planets. On the other hand, we considered as important to check the possible TTVs in the case of our sample, because precise \textit{CHEOPS} data can uncover such a variation with higher probability.       

\subsection{Search for long-term TDVs -- signs of orbital precession}
\label{tdv}

Long-term variations in transit duration, caused by orbital precession, can be detected more easily than variations in mid-transit times, because such a long-term TDV is a linear function of time, thus longer time base-line, greater difference in transit duration \citep{Pal1}. \textit{CHEOPS} observations of KELT-17, KELT-19A, and KELT-21 cover a time-baseline of about two months (see Table \ref{cheopsobslog}), which is a very short interval from this point of view. Another problem is that due to interruptions in \textit{CHEOPS} observations (see Sect. \ref{phot}) the ingress and/or the egress part of the transit light curve can be missed, thus it is not possible to search for TDVs only using the obtained \textit{CHEOPS} data, i.e., from transit to transit. Therefore, we used the \textit{TESS} data of these objects and the available literature data to increase the time-baseline. 

\begin{figure*}
\centering
\centerline{
\includegraphics[width=\columnwidth]{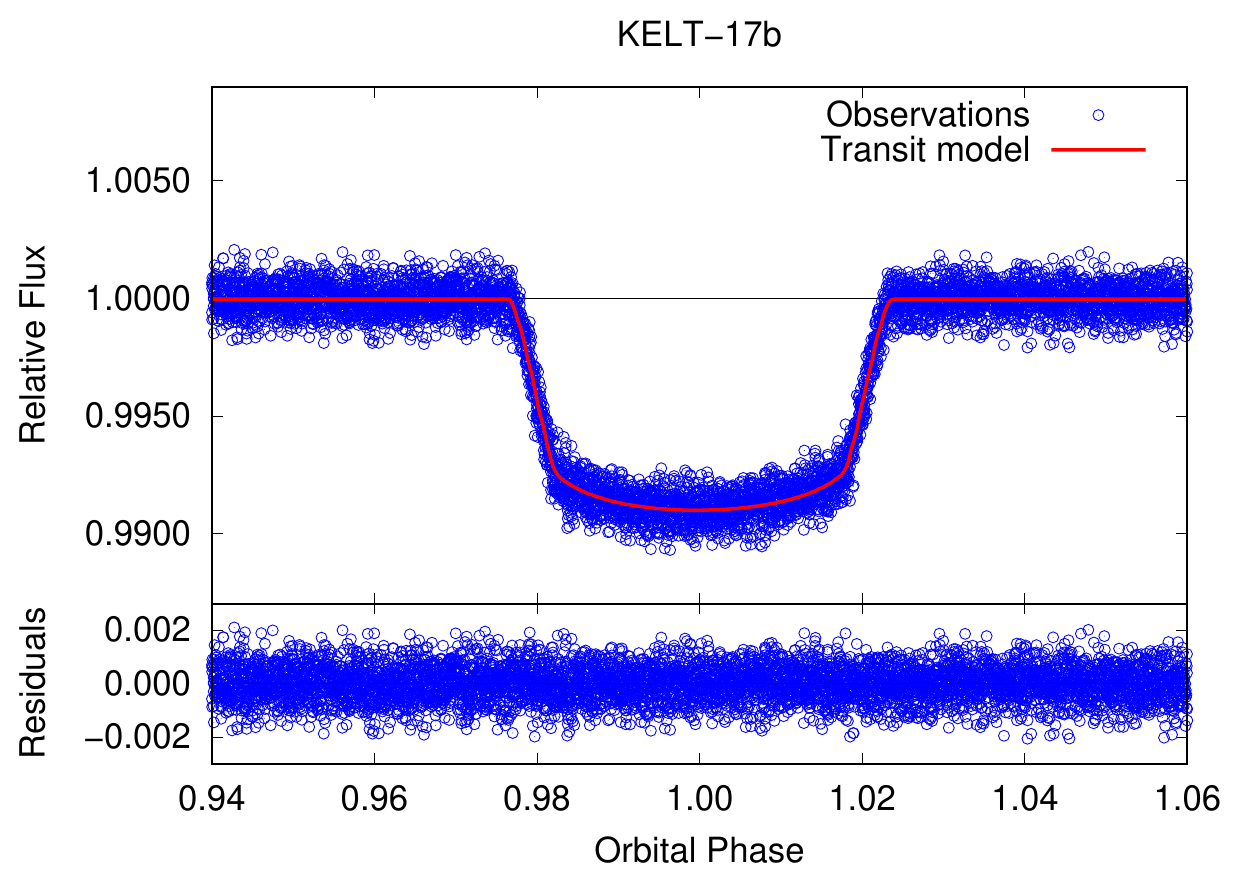}
\includegraphics[width=\columnwidth]{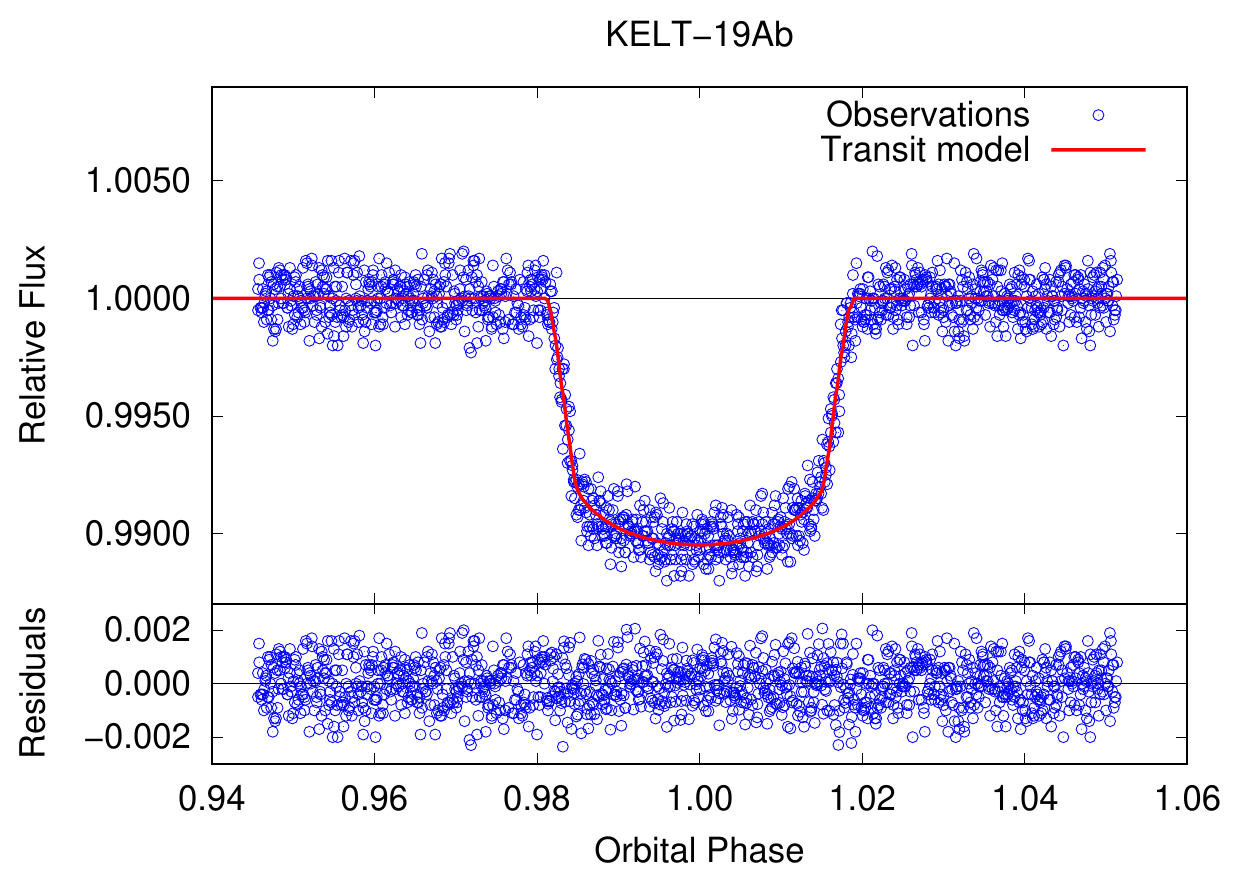}}
\caption{Phase-folded \textit{TESS} transit light curves of KELT-17b (left-hand panel) and KELT-19Ab (right-hand panel), overplotted with the best-fitting \rmf models. Residuals are also shown (bottom panels). During this modeling procedure all \textit{TESS} data per object were fitted and the time-baseline of observations was extended for the purpose of search for TDVs.}
\label{kelt1719fittedtesslcrmf} 
\end{figure*}

\begin{figure}
\includegraphics[width=\columnwidth]{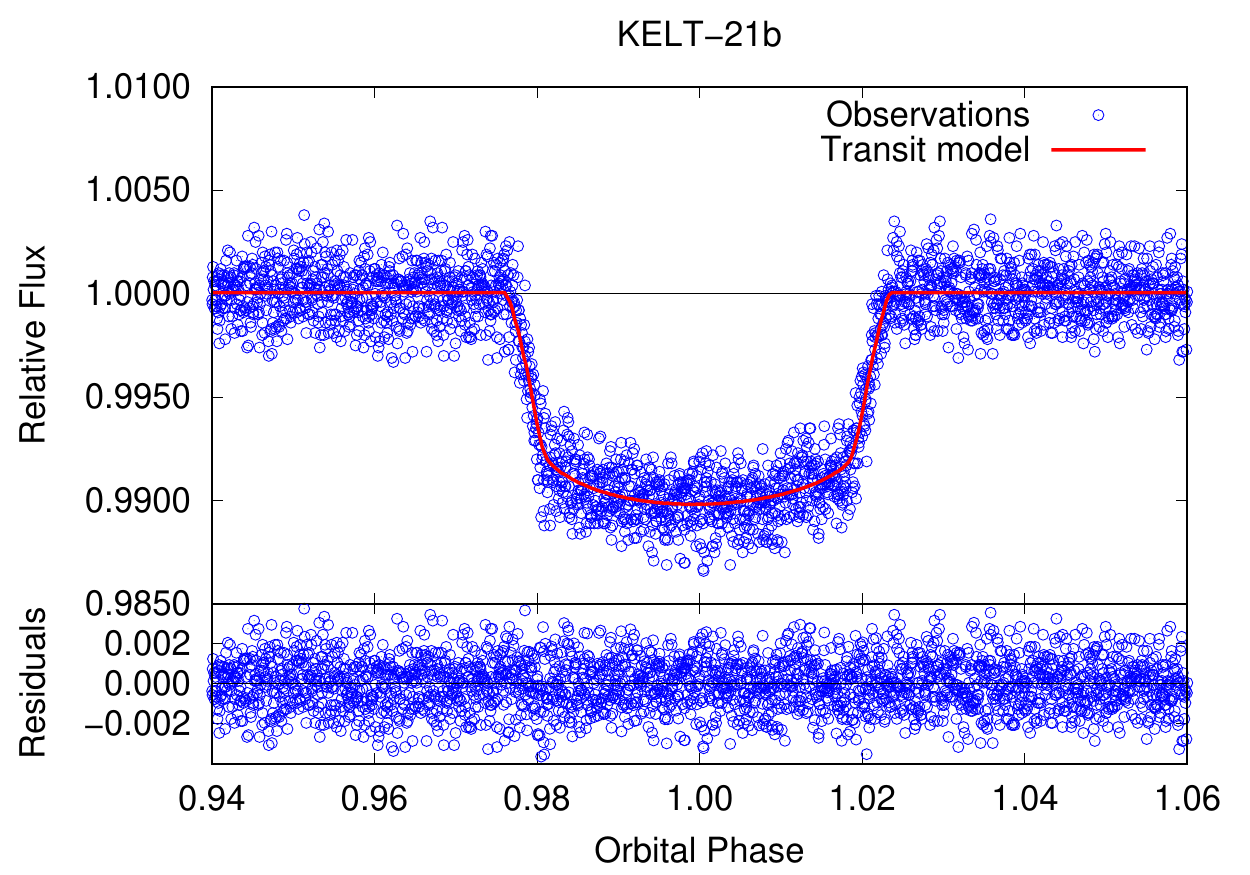}
\caption{As in Fig. \ref{kelt1719fittedtesslcrmf}, but for KELT-21b.}
\label{kelt21fittedtesslcrmf} 
\end{figure}

KELT-17b was observed with \textit{TESS} during three Sectors Nos. 44, 45, and 46 from 2021-10-12 to 2021-12-30. KELT-19Ab was observed in Sector No. 7 from 2019-01-07 to 2019-02-02. KELT-21b was observed in Sector No. 41 from 2021-07-23 to 2021-08-20. The data were downloaded from the Mikulski Archive for Space Telescopes\footnote{See \url{https://mast.stsci.edu/portal/Mashup/Clients/Mast/Portal.html}.} in the form of Simple Aperture Photometry (SAP) fluxes, see Table \ref{TESSobslog} for the \textit{TESS} observational log. These data were obtained from 2-min integrations, but in comparison with Pre-search Data Conditioning Simple Aperture Photometry (PDCSAP) fluxes, long-term trends were not removed. The downloaded SAP fluxes were detrended using our pipeline, as follows. The SAP fluxes were first normalized to unity. During the next step \textit{TESS} data were cut into segments, each covering one orbital period. Each segment of the data was fitted with a linear function. During the fitting procedure the part of the data covering the transit was excluded from the fit. Consequently, the linear trend was removed from each chunk of data (including the transit data). This detrending method can effectively remove the long-term variability (mainly variability of the host star due to spots and rotation) while it does not introduce any nonlinear trend to the data, see e.g., \citet{Garai3}. Outliers were cleaned using a $3\sigma$ clipping, where $\sigma$ is the standard deviation of the light curve. Since \textit{TESS} uses as time-stamps Barycentric \textit{TESS} Julian Date (i.e., $\mathrm{BJD}_\mathrm{TDB} - 2~457~000.0$), during the next step we converted all \textit{TESS} time-stamps to $\mathrm{BJD}_\mathrm{TDB}$. We analyzed the detrended \textit{TESS} photometry data using the \rmf code, described in Sect. \ref{spinorb}. During this analysis procedure we used the same free and fixed parameters as we presented in Table \ref{rmfres}. We applied the four-parameter limb-darkening model with the critical foreshortening angle approach \citep{Claret1}. The coefficients were calculated for the \textit{TESS} passband using the same spherical {\tt{PHOENIX-COND}} models as in \citet{Claret1}. The applied limb-darkening coefficients were first linearly interpolated from the calculated table and then were used identically, as it is described in Sect. \ref{spinorb}. Since we used \textit{TESS} SAP fluxes, which are not corrected by the dilution factor, we used the CROWDSAP\footnote{CROWDSAP is a keyword on the header of the FITS files containing the light curves. It represents the ratio of the target flux to the total flux in the \textit{TESS} aperture.} crowding metric value to determine the $l_3$ parameter for the \textit{TESS} aperture. In the case of KELT-17b this gives $l_{3,\mathrm{\textit{TESS}}} = 0.0009$, for KELT-19Ab $l_{3,\mathrm{\textit{TESS}}} = 0.0150$, and in the case of KELT-21b $l_{3,\mathrm{\textit{TESS}}} = 0.0636$. The uncertainties in the fitted parameters were derived based on the covariance matrix method. We present the best-fitting parameters of KELT-17, KELT-19A, and KELT-21 systems in Table \ref{rmfres2}. The corresponding phase-folded transit light curves, overplotted with the best-fitting \rmf models are depicted in Figs. \ref{kelt1719fittedtesslcrmf} and \ref{kelt21fittedtesslcrmf}.

\begin{table}
\centering
\caption{Log of \textit{TESS} observations of KELT17b, KELT-19Ab and KELT-21b used in our analysis. Table shows time interval of observations, number of observed transits, and number of data points obtained from the \textit{TESS} database.}
\label{TESSobslog}
\begin{tabular}{cccc}
\hline
\hline
Target & Time interval of observations & Transits & Data points\\
\hline
\hline
KELT-17         & 2021-10-12 -- 2021-12-30      & 22    & 47~411\\ 
KELT-19A 	& 2019-01-07 -- 2019-02-02	& 4	& 16~362\\
KELT-21 	& 2021-07-23 -- 2021-08-20	& 8	& 18~322\\
\hline
\hline
\end{tabular}
\end{table}

\begin{table*}
\centering
\caption{An overview of the \rmf best-fitting parameters of the exoplanets KELT-17b, KELT-19Ab, and KELT-21b obtained from the \textit{TESS} photometry and using the gravity-darkening approach of \citet{Espinosa1}. Only the fitted parameters and the $l_3$ parameter values are presented here. Other parameters were fixed as we presented in Table \ref{rmfres}. Notes: $^1$The closest mid-transit time to the midpoint of the \textit{TESS} dataset.}
\label{rmfres2}
\begin{tabular}{llll}
\hline
\hline
Parameter [unit] & KELT-17b & KELT-19Ab & KELT-21b\\
\hline
\hline
$T_\mathrm{c}$ [$\mathrm{BJD}_\mathrm{TDB}$]$^1$ 	& $2~459~536.27682 \pm 0.00013$ 	& $2~458~503.35923 \pm 0.00017$ 	& $2~459~434.69341 \pm 0.00019$\\
$i$ [deg] 						& $84.537 \pm 0.061$ 			& $87.65 \pm 0.36$ 			& $86.95 \pm 0.18$\\
$R_\mathrm{s}/a$ 					& $0.16141 \pm 0.00063$ 		& $0.1148 \pm 0.0023$ 			& $0.1452 \pm 0.0013$\\
\rprs 							& $0.09166 \pm 0.00011$ 		& $0.09550 \pm 0.00030$ 		& $0.09960 \pm 0.00040$\\
$l_\mathrm{norm}$ 					& $1.0000740 \pm 0.0000010$ 		& $1.0000750 \pm 0.0000010$ 		& $1.0000740 \pm 0.0000010$\\
$l_3$ (fixed) 						& $0.0009$ 				& $0.0150$ 				& $0.0636$\\
\hline
\hline
\end{tabular}
\end{table*}

Based on the mid-transit time values $T_\mathrm{c}$, presented in the Tables \ref{photometryres} and \ref{rmfres2}, we can clearly see that the time-baseline was extended significantly using the \textit{TESS} database. In the case of KELT-17b, the \textit{TESS} dataset follows the \textit{CHEOPS} dataset. The closest mid-transit time to the midpoint of \textit{TESS} observations is $T_\mathrm{c,\textit{TESS}}= 2~459~536.27682 \pm 0.00013~\mathrm{BJD}_\mathrm{TDB}$, while for the \textit{CHEOPS} dataset it is $T_\mathrm{c,\textit{CHEOPS}}= 2~459~215.937950 \pm 0.000086~\mathrm{BJD}_\mathrm{TDB}$. The time-baseline of these two datasets is about one year, but if we add the literature data \citep{Zhou1} to the dataset, based on the mid-transit time of $T_\mathrm{c,Z2016} = 2~457~226.14219 \pm 0.00033~\mathrm{BJD}_\mathrm{TDB}$ we can extend the time-baseline by about 1989 days. In the case of KELT-19Ab, the \textit{TESS} dataset precedes \textit{CHEOPS} observations. The mid-transit time of \textit{TESS} data is $T_\mathrm{c,\textit{TESS}}= 2~458~503.35923 \pm 0.00017~\mathrm{BJD}_\mathrm{TDB}$, while for the \textit{CHEOPS} dataset it is $T_\mathrm{c,\textit{CHEOPS}}= 2~459~218.17799 \pm 0.00013~\mathrm{BJD}_\mathrm{TDB}$. The time-baseline of these two datasets is about two years. Using the literature data presented by \citet{Siverd1}, we could extend this time-baseline by about 1222 days -- this is the difference between the mid-transit time $T_\mathrm{c,S2018} = 2~457~281.24953 \pm 0.00036~\mathrm{BJD}_\mathrm{TDB}$, presented by these authors and the mid-transit time derived based on the \textit{TESS} data. In the case of KELT-21b, the \textit{TESS} dataset follows the \textit{CHEOPS} dataset. Similarly, we can write the mid-transit time values of $T_\mathrm{c,\textit{CHEOPS}}= 2~459~055.352380 \pm 0.000086~\mathrm{BJD}_\mathrm{TDB}$ and $T_\mathrm{c,\textit{TESS}}= 2~459~434.69341 \pm 0.00019~\mathrm{BJD}_\mathrm{TDB}$, which gives about one year difference, but if we also include the data presented in the literature \citep{Johnson1}, we can extend this time-baseline by about 1672 days ($T_\mathrm{c,J2018} = 2~457~382.640727 \pm 0.00041~\mathrm{BJD}_\mathrm{TDB}$). 

\begin{figure*}
\centering
\centerline{
\includegraphics[width=\columnwidth]{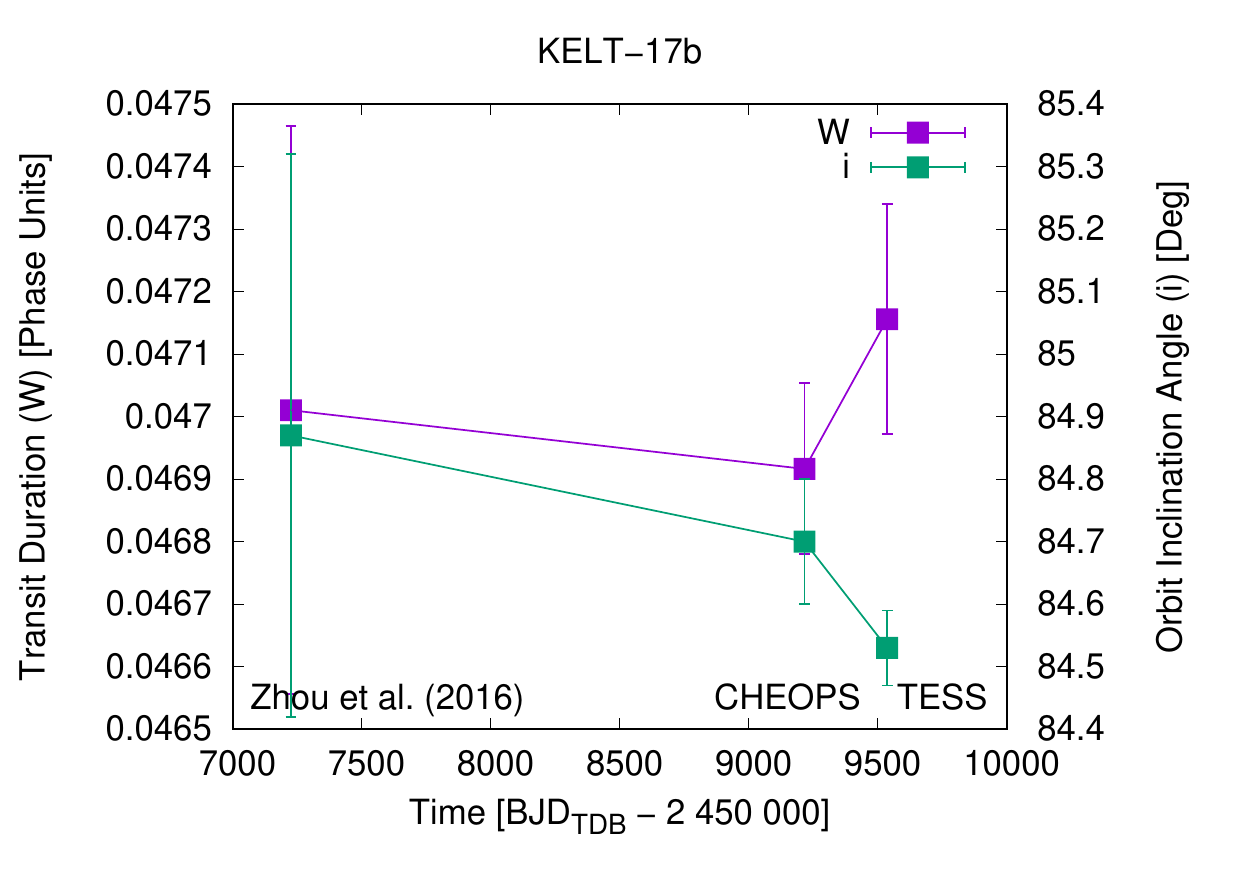}
\includegraphics[width=\columnwidth]{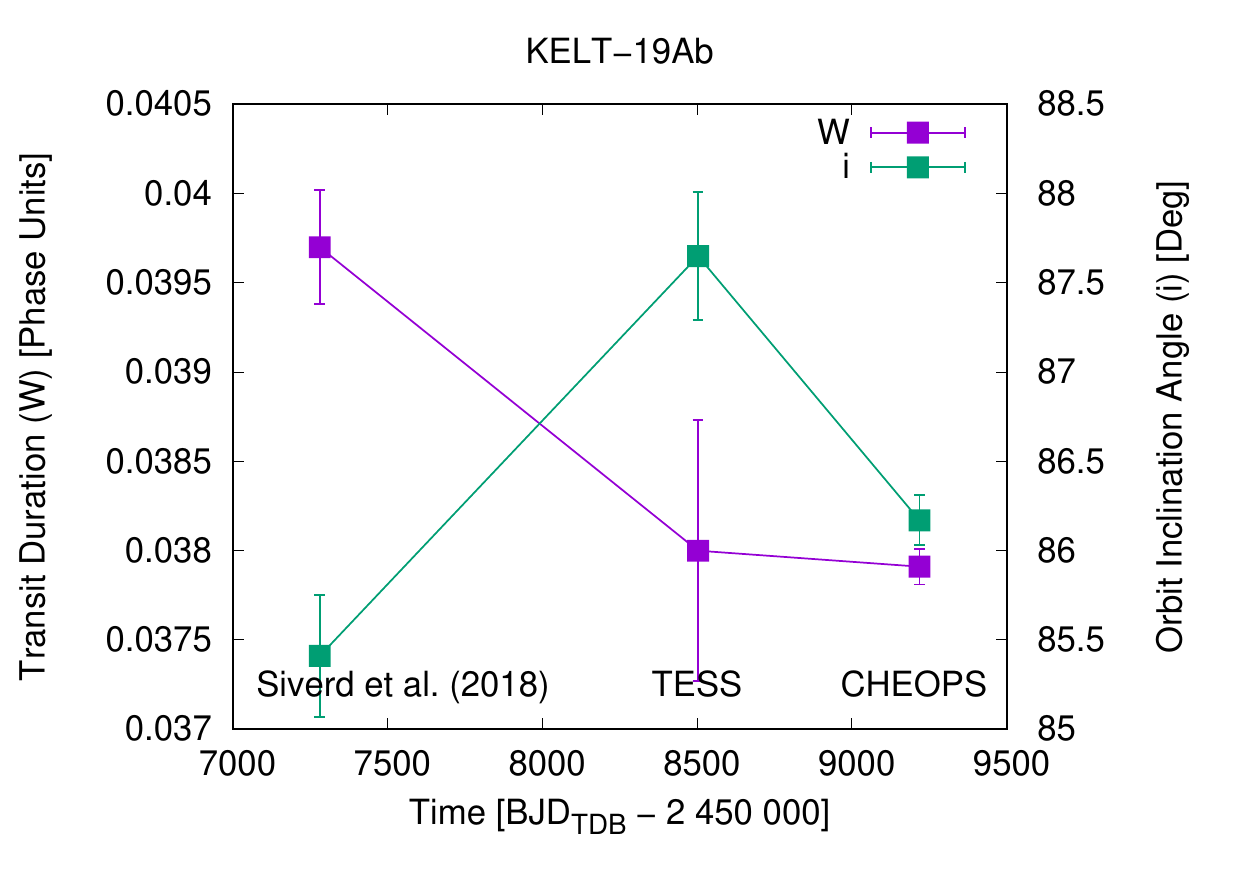}}
\caption{Transit durations and orbit inclination angles of KELT-17b (left-hand panel) and KELT-19Ab (right-hand panel) in different seasons, obtained based on literature data, \textit{CHEOPS}, and \textit{TESS} observations. For more details see the text of Sect. \ref{tdv}.}
\label{kelt1719tdv} 
\end{figure*}

\begin{figure}
\includegraphics[width=\columnwidth]{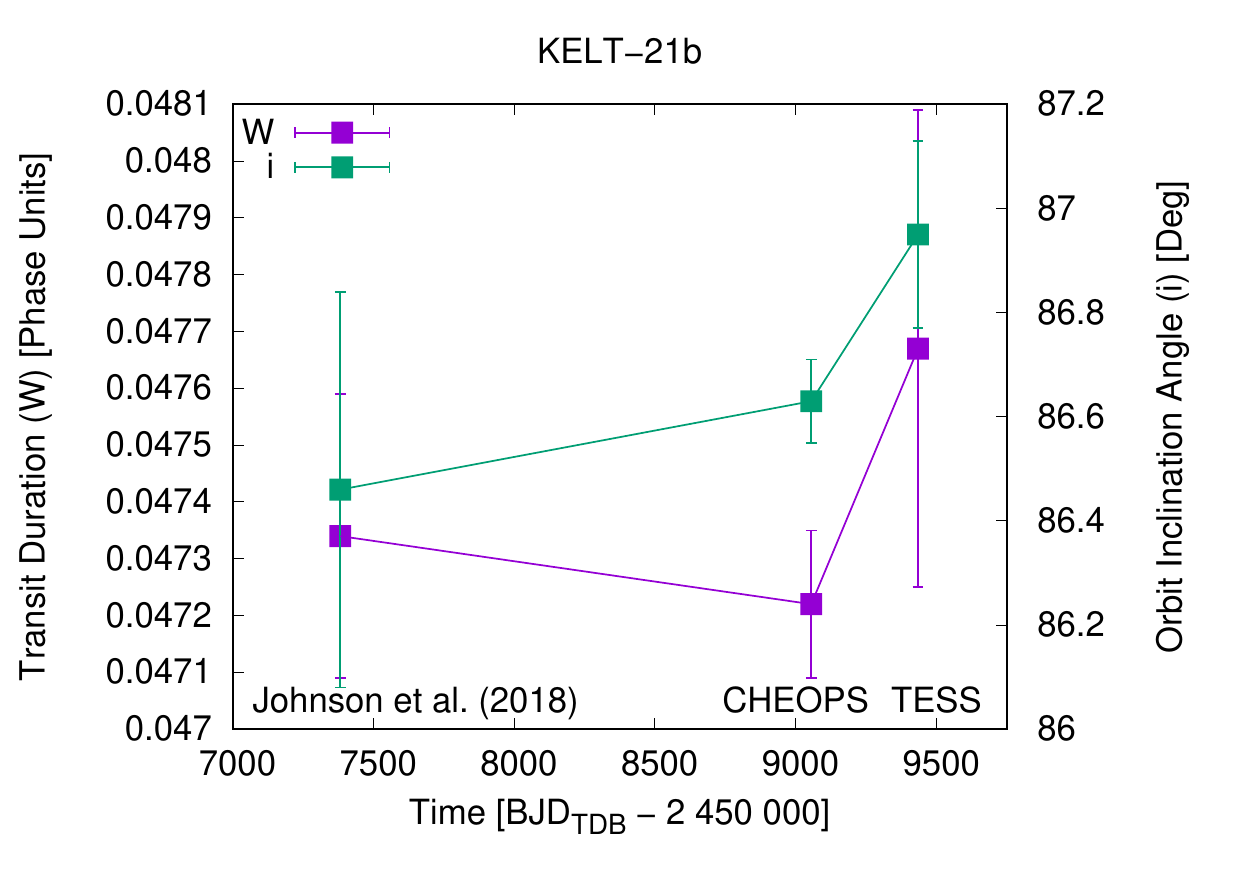}
\caption{As in Fig. \ref{kelt1719tdv}, but for KELT-21b.}
\label{kelt21tdv} 
\end{figure}

The \texttt{pycheops} software uses the transit width $W$ as a free parameter, which corresponds to the transit duration of \textit{CHEOPS} visits in phase units. In the cited literature we can also find the tabulated transit duration of the exoplanets in days, which we can easily convert to the phase units. The most problematic are the \textit{TESS} data fitted with the \rmf code. This software does not use such a free parameter, therefore we directly measured the transit durations in the plots presented in Figs. \ref{kelt1719fittedtesslcrmf} and \ref{kelt21fittedtesslcrmf}, which are also expressed in phase units. The uncertainties in these cases follow from the uncertainties of $R_\mathrm{s}/a$ and $i$ (see Table \ref{rmfres2}), since these parameters affect the transit duration most significantly. In this way we could compare the transit durations of the exoplanets coming from three seasons of observation, which is enough to uncover possible long-term TDVs -- signs of orbital precession. Simultaneously, we also checked the orbit inclination angle values, because the possible orbital precession should be visible in this parameter, as well \citep{Szabo2, Szabo3}. We note that the orbit inclination angle change rate $\mathrm{d}i/\mathrm{d}t$ parameter used by \citet{Szabo3} is applicable only if long-term consecutive observations exist, e.g., \textit{Kepler} observations, but this is not the case, therefore we did not apply this parameter during the analysis. The impact parameters $b$ obtained using the \texttt{pycheops} software were converted to the orbit inclination angle values $i$. In such a combination of parameters, i.e., $W$ and $i$, we can assume that if a long-term TDV is caused by orbital precession, a change in the parameter $W$ is correlated with the change in parameter $i$. This means, for example, that if $W$ is increasing, $i$ is also increasing and vice versa. In the case of $W$ and $b$ the trend should be anticorrelated, see \citet{Szabo2}. This should be taken into consideration during the interpretation of the obtained results, which are plotted in Figs. \ref{kelt1719tdv} and \ref{kelt21tdv}.            

Here we can see the transit durations $W$ and the orbit inclination angles $i$ in three different seasons. Since \textit{CHEOPS} data were analyzed with the \rmf code, as well as with the \texttt{pycheops} software, we present here only the latter solution as finally adopted results. In the case of KELT-17b (Fig. \ref{kelt1719tdv} left-hand panel) we can see that the \textit{CHEOPS} data reveal slightly smaller values of $W$ and $i$ in comparison with the discoverers, but due to the large uncertainties in the discovery values this difference is inconclusive. Using \textit{TESS} observations we obtained the $W$ and $i$ parameter values within a $3\sigma$ agreement with the \textit{CHEOPS} results, which means that the difference is not significant. Moreover, the parameter trends seem to be anticorrelated between the \textit{CHEOPS} and \textit{TESS} observations, which is against the precession-based long-term TDV, as we discussed this in the previous paragraph. KELT-19Ab is also an interesting object, since using both (\textit{TESS} and \textit{CHEOPS}) datasets we found the transit duration shorter and the orbit inclination angle value larger in comparison with the discovery values, see Fig. \ref{kelt1719tdv} right-hand panel. The difference between certain parameter values is significant, exceeding the $3\sigma$ difference, which indicates the robustness of the detection. If we took a look again at Fig. \ref{kelt1719tdv} right-hand panel, we can see that the trend in $W$ and $i$ is anticorrelated, thus the detection of the shorter transit duration compared to the discovery paper cannot be due to orbital precession. Rather, the smaller $W$ is a consequence of a smaller planet size, as we concluded in Sect. \ref{k19Abcheops}. In the case of KELT-21b (see Fig. \ref{kelt21tdv}), using \textit{CHEOPS} and \textit{TESS} observations we found the orbit inclination angle larger in comparison with the discovery value and the transit duration is also longer based on the \textit{TESS} measurements. Although the differences are not significant due to the larger uncertainties in the detection, we can see that the trend in $W$ and $i$ seems to be correlated, thus this object could be interesting from the viewpoint of long-term TDV and orbital precession in the future. More observations are needed in this case.                              

\subsection{Search for TTVs -- signs of additional substellar objects in the systems}
\label{ttv}

To uncover the possible TTVs, which can indicate additional substellar objects, i.e., planets or brown dwarfs in the systems, we constructed the observed-minus-calculated (O-C) diagram for mid-transit times. Similarly as in Sect. \ref{tdv}, we used not only the \textit{CHEOPS} observations, because of the following reasons. The interruptions occurring in \textit{CHEOPS} visits represent a disadvantage from this viewpoint, although the data are precise. If the ingress and/or the egress part of the transit is missing, this can significantly decrease the precision of the O-C value. Another reason for use of additional observations is that more O-C data can easier uncover any periodicity coming from perturbations of an additional object in the system. Therefore, besides the \textit{CHEOPS} data collected within this project, we also used the available \textit{TESS} data and literature data. \textit{CHEOPS} data were treated using the \texttt{pycheops} software as it is described in the second part of Sect. \ref{jointanalysis}. During this procedure we obtained four fitted $\Delta T_{0,\mathrm{n}}$ values, i.e., one for each visit. Since $\Delta T_{0,\mathrm{n}}$ values correspond to the wanted O-C values of mid-transit times, we could easily get the O-C diagram of \textit{CHEOPS} observations. Such a diagram is generated automatically by the \texttt{pycheops} software and it enabled us to first-look check these O-C data. On the other hand, we needed the 'O' times of individual transits with uncertainties, i.e., the $T_\mathrm{0,n}$ values, since this is the input for the \texttt{OCFIT} code, described and applied later in this section. Therefore, we used Eq. \ref{ttvpycheops} to get the observed mid-transit times of individual \textit{CHEOPS} transits. In the next step we used the \textit{TESS} data, already processed in Sect. \ref{tdv}. To obtain the 'O' times of the mid-transits we modeled each \textit{TESS} transit event individually using the \rmf code. We fixed every parameter to its best value from the joint \texttt{pycheops} model (see Table \ref{photometryres}) except for two parameters: the mid-transit time and the light-curve normalization factor of the given transit. We summarize the \textit{CHEOPS} and \textit{TESS} observed mid-transit times in Table \ref{otimes}. Finally, to extend the time-baseline of \textit{CHEOPS} and \textit{TESS} observations and to increase the amount of data-points, we also used the already published 'O' times, presented by \citet{Zhou1}, \citet{Siverd1}, and \citet{Johnson1}.

\begin{table}
\centering
\caption{The list of the observed ('O') mid-transit times of KELT-17b, KELT-19Ab, and KELT-21b, derived in this work using the joint \texttt{pycheops} model parameter values (see Table \ref{photometryres}). The epoch $E$ is calculated according to the $T_\mathrm{c}$ parameter values, presented by \citet{Zhou1}, \citet{Siverd1}, and \citet{Johnson1}.}
\label{otimes}
\begin{tabular}{cccc}
\hline
\hline
$E$ & 'O' times [$\mathrm{BJD}_\mathrm{TDB}$] & $\pm 1\sigma$ [d] & Source\\
\hline
\hline
\multicolumn{4}{c}{KELT-17b}\\
639 & 2459194.37646 & 0.00041 & \textit{CHEOPS}\\
641 & 2459200.53715 & 0.00016 & \textit{CHEOPS}\\ 
642 & 2459203.61727 & 0.00015 & \textit{CHEOPS}\\
661 & 2459262.14059 & 0.00016 & \textit{CHEOPS}\\
739 & 2459502.39443 & 0.00032 & \textit{TESS}\\    
740 & 2459505.47552 & 0.00028 & \textit{TESS}\\  
741 & 2459508.55519 & 0.00030 & \textit{TESS}\\  
742 & 2459511.63552 & 0.00032 & \textit{TESS}\\  
743 & 2459514.71578 & 0.00031 & \textit{TESS}\\  
744 & 2459517.79631 & 0.00034 & \textit{TESS}\\  
745 & 2459520.87523 & 0.00030 & \textit{TESS}\\  
746 & 2459523.95613 & 0.00030 & \textit{TESS}\\  
747 & 2459527.03613 & 0.00029 & \textit{TESS}\\  
748 & 2459530.11631 & 0.00027 & \textit{TESS}\\  
749 & 2459533.19632 & 0.00030 & \textit{TESS}\\  
750 & 2459536.27697 & 0.00030 & \textit{TESS}\\  
752 & 2459542.43692 & 0.00029 & \textit{TESS}\\  
753 & 2459545.51797 & 0.00029 & \textit{TESS}\\  
754 & 2459548.59758 & 0.00030 & \textit{TESS}\\  
756 & 2459554.75829 & 0.00031 & \textit{TESS}\\  
757 & 2459557.83755 & 0.00032 & \textit{TESS}\\  
758 & 2459560.91815 & 0.00027 & \textit{TESS}\\  
759 & 2459563.99834 & 0.00032 & \textit{TESS}\\  
761 & 2459570.15903 & 0.00028 & \textit{TESS}\\  
762 & 2459573.23897 & 0.00028 & \textit{TESS}\\  
763 & 2459576.31897 & 0.00030 & \textit{TESS}\\  
\hline
\multicolumn{4}{c}{KELT-19Ab}\\
263 & 2458494.13568 & 0.00047 & \textit{TESS}\\
264 & 2458498.74760 & 0.00038 & \textit{TESS}\\
266 & 2458507.97115 & 0.00048 & \textit{TESS}\\
267 & 2458512.58297 & 0.00043 & \textit{TESS}\\
412 & 2459181.28423 & 0.00019 & \textit{CHEOPS}\\
419 & 2459213.56631 & 0.00023 & \textit{CHEOPS}\\
425 & 2459241.23668 & 0.00028 & \textit{CHEOPS}\\
426 & 2459245.84811 & 0.00031 & \textit{CHEOPS}\\
\hline
\multicolumn{4}{c}{KELT-21b}\\
457 & 2459033.67650 & 0.00032 & \textit{CHEOPS}\\
462 & 2459051.74102 & 0.00071 & \textit{CHEOPS}\\
463 & 2459055.35200 & 0.00021 & \textit{CHEOPS}\\
472 & 2459087.86668 & 0.00046 & \textit{CHEOPS}\\
564 & 2459420.24231 & 0.00058 & \textit{TESS}\\
565 & 2459423.85549 & 0.00054 & \textit{TESS}\\
566 & 2459427.46758 & 0.00060 & \textit{TESS}\\
567 & 2459431.08095 & 0.00059 & \textit{TESS}\\
568 & 2459434.69341 & 0.00053 & \textit{TESS}\\
569 & 2459438.30580 & 0.00052 & \textit{TESS}\\
570 & 2459441.91831 & 0.00050 & \textit{TESS}\\
571 & 2459445.53197 & 0.00050 & \textit{TESS}\\
\hline
\hline
\end{tabular}
\end{table}

\begin{figure*}
\centering
\centerline{
\includegraphics[width=\columnwidth]{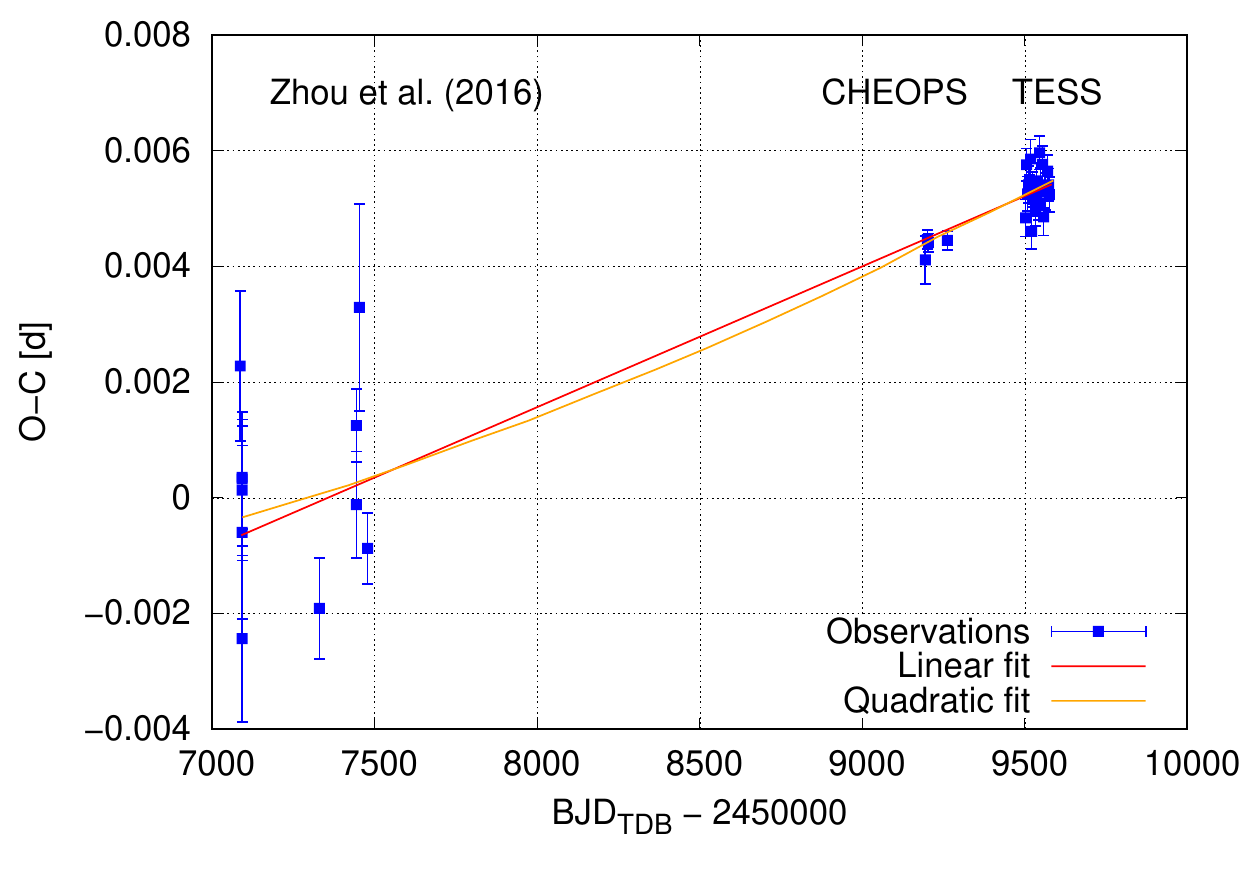}
\includegraphics[width=\columnwidth]{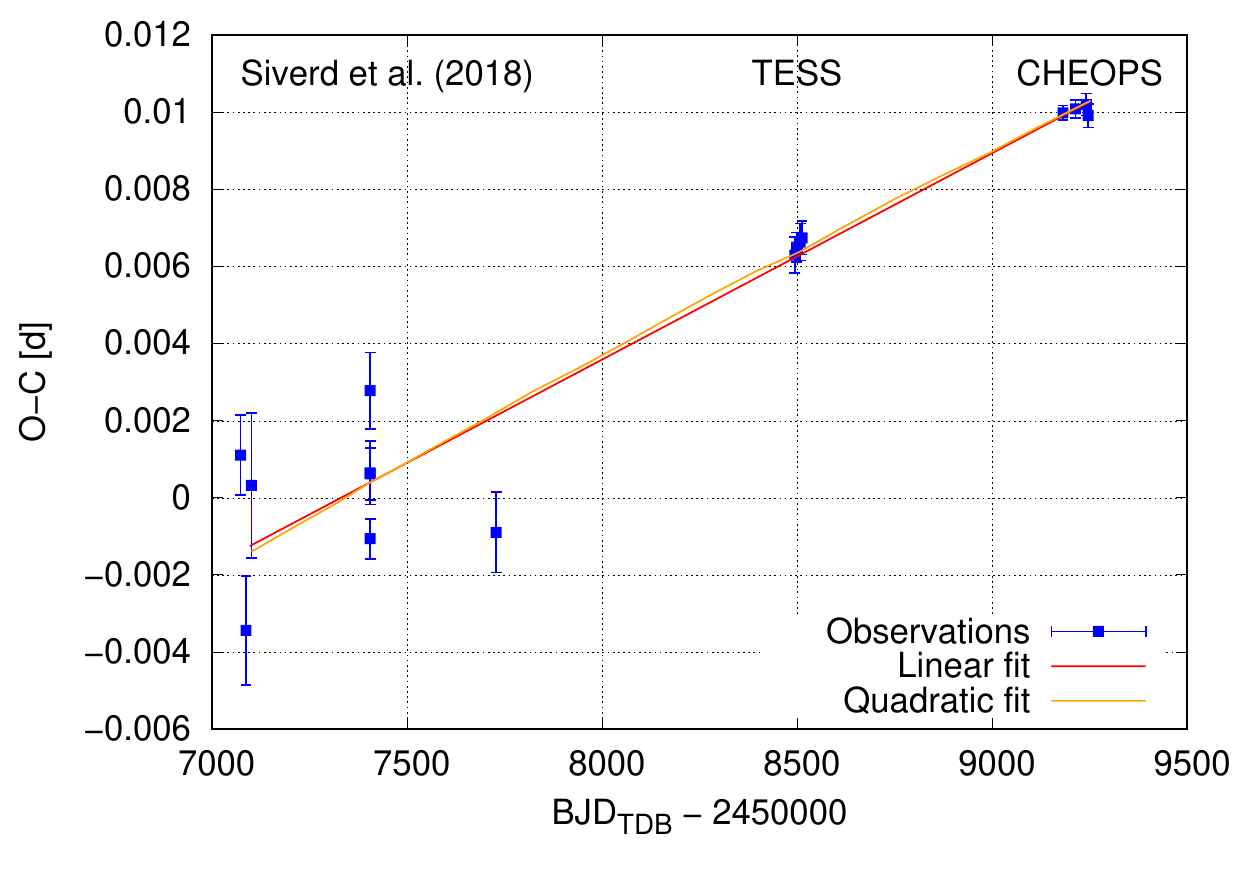}}
\caption{Observed-minus-calculated (O-C) diagrams of KELT-17b (left-hand panel) and KELT-19Ab (right-hand panel) mid-transit times, obtained based on literature data, \textit{CHEOPS}, and \textit{TESS} observations. For more details see the text of Sect. \ref{ttv}.}
\label{kelt1719ttv} 
\end{figure*} 

\begin{figure}
\includegraphics[width=\columnwidth]{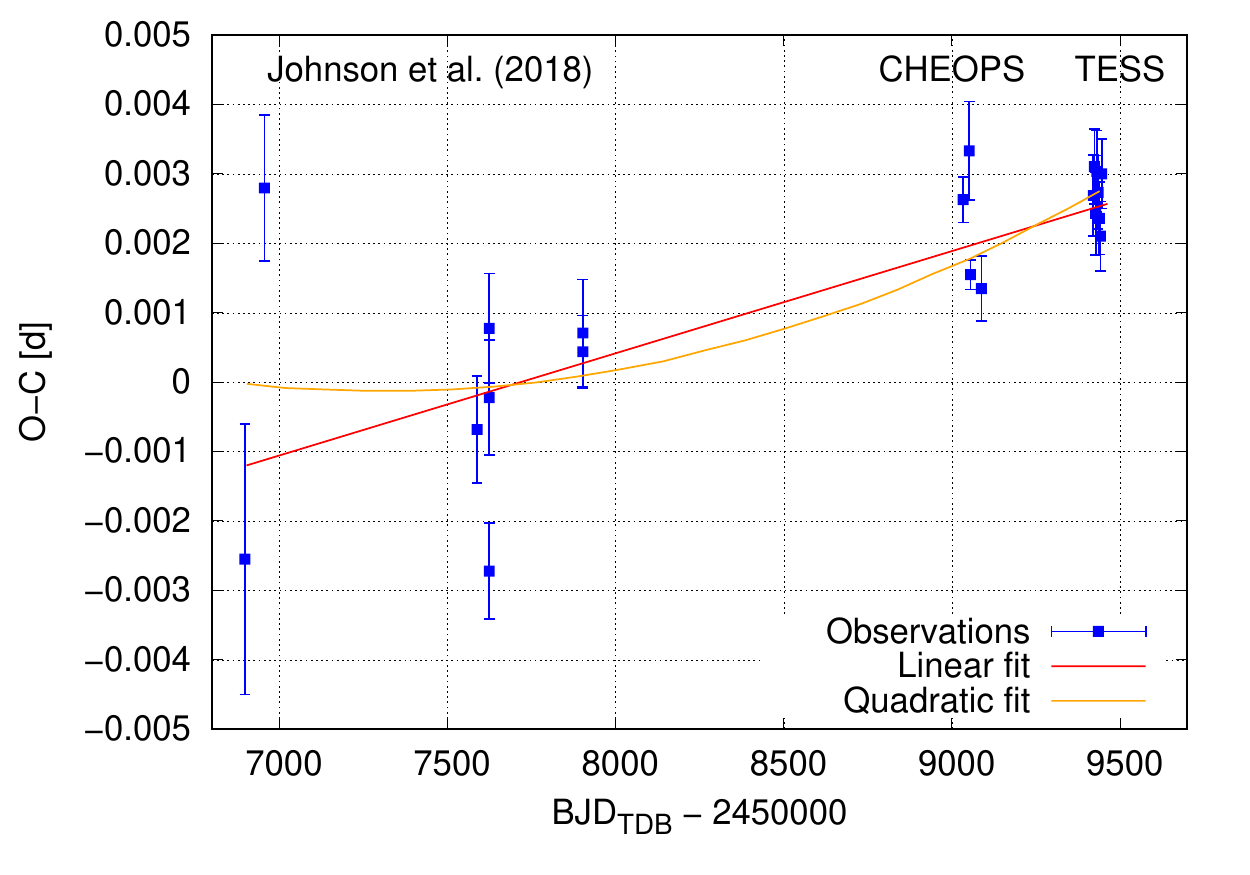}
\caption{As in Fig. \ref{kelt1719ttv}, but for KELT-21b.}
\label{kelt21ttv} 
\end{figure}

\begin{table}
\centering
\caption{Linear ephemeris of KELT-17b, KELT-19Ab, and KELT-21b, obtained based on the whole O-C data-set of mid-transit times. Here we improved further the orbital period of the planets compared to the joint \texttt{pycheops} solutions (see and compare the appropriate parameter values with Table \ref{photometryres}), thus the final $P_\mathrm{orb}$ values are presented in this table.}  
\label{keltslineareph}
\begin{tabular}{lcc}
\hline
\hline
Parameter [unit] & Value & $\pm 1\sigma$\\
\hline
\hline
\multicolumn{3}{c}{KELT-17b}\\
$T_\mathrm{c}$ [$\mathrm{BJD}_\mathrm{TDB}$] & $2~457~226.14186$ & $0.00041$\\  
$P_\mathrm{orb}$ [d] & $3.08017988$ & $0.00000058$\\
\hline
\multicolumn{3}{c}{KELT-19Ab}\\
$T_\mathrm{c}$ [$\mathrm{BJD}_\mathrm{TDB}$] & $2~457~281.24924$ & $0.00069$\\
$P_\mathrm{orb}$ [d] & $4.6117352$ & $0.0000018$\\
\hline
\multicolumn{3}{c}{KELT-21b}\\
$T_\mathrm{c}$ [$\mathrm{BJD}_\mathrm{TDB}$] & $2~457~382.64023$ & $0.00083$\\
$P_\mathrm{orb}$ [d] & $3.6127693$ & $0.0000017$\\
\hline 
\hline
\end{tabular}
\end{table}

To calculate the 'C' times of the individual transit events and to construct the O-C diagram of mid-transit times we applied the \texttt{OCFIT}\footnote{See \url{https://github.com/pavolgaj/OCFit}.} code \citep{Gajdos2}. To plot the O-C diagram of the mid-transit times it requires 'O' times with uncertainties, the mid-transit time $T_\mathrm{c}$ and the orbital period $P_\mathrm{orb}$ of the planet. Then it uses the linear ephemeris formula as:

\begin{equation}
\label{lineph}
T_\mathrm{c,n} = T_\mathrm{c} + P_\mathrm{orb} \times E
\end{equation} 

\noindent to construct the O-C diagram, were $T_\mathrm{c,n}$ corresponds to the 'C' value of the $n$-th transit, and $E$ is the epoch of observation, i.e., the number of the orbital cycle. For $P_\mathrm{orb}$ we used the best-fitting joint \texttt{pycheops} model parameters (see Table \ref{photometryres}), but for $T_\mathrm{c}$ we chose the literature values presented by \citet{Zhou1}, \citet{Siverd1}, and \citet{Johnson1}, because the derived \textit{CHEOPS}-based $T_\mathrm{c}$ parameter values were located at the end/middle of the joint (literature and this work) datasets. It is better to choose the $T_\mathrm{c}$ parameter from the beginning of the whole dataset. The obtained O-C diagrams of the planets are depicted in Figs. \ref{kelt1719ttv} and \ref{kelt21ttv}. We can see that the O-C data of mid-transit times do not show periodic features, which means that there is no evidence for a third body in the planetary systems. To final check the possibility of additional objects in the systems we fitted the O-C datasets of mid-transit times. As first we applied a linear function using the \texttt{OCFIT} package \texttt{FitLinear}. The free parameters of the linear model are the mid-transit time $T_\mathrm{c}$ and the orbital period $P_\mathrm{orb}$. Subsequently, the O-C data were fitted with a quadratic function within the \texttt{OCFIT} package \texttt{FitQuad}. The free parameters of the quadratic model are the mid-transit time $T_\mathrm{c}$, the orbital period $P_\mathrm{orb}$, and the quadratic coefficient $Q$, see \citet{Garai3} for more information about this parameter. The uncertainties in the fitted parameters of $P_\mathrm{orb}$, $T_\mathrm{c}$, and $Q$ were derived within the \texttt{OCFIT} packages \texttt{FitLinear} and \texttt{FitQuad}, applying the covariance matrix method. These fits are also visible in Figs. \ref{kelt1719ttv} and \ref{kelt21ttv}. The quality of the linear and quadratic fits was expressed as Bayesian Information Criterion ($BIC$), which is defined as:

\begin{equation}
BIC = \chi^2 + k \ln N,
\label{bic}
\end{equation}

\noindent where $k$ is the number of free parameters of the model and $N$ is the number of data-points.

In the case of KELT-17b the observations cover a time-baseline of about 2400 days, but the literature data have significant scatter compared to the \textit{CHEOPS} and \textit{TESS} data. There is no significant difference between the Bayesian Information Criterions when we apply linear, or quadratic fit ($BIC_\mathrm{lin} = 59.0$ and $BIC_\mathrm{quad} = 60.3$, respectively), therefore the quadratic fit is not justified here. The O-C values obtained by the discoverers allowed to exclude periodic TTVs with the semi-amplitude of about 5 min. If we take into consideration \textit{CHEOPS} and \textit{TESS} observations, we can put more constraints on this upper limit with the value of about 3 min ($3\sigma$ upper limit). Based on the linear fit we obtained a new linear ephemeris of KELT-17b, which is presented in Table \ref{keltslineareph}. For KELT-19Ab we collected transit observations covering a time baseline of about 2000 days, including the literature data. The O-C data have a well defined and clearly visible linear trend, see Fig. \ref{kelt1719ttv} (right-hand panel). There is no significant difference between the linear and quadratic fit, the Bayesian Information Criterions are very similar, i.e., $BIC_\mathrm{lin} = 39.7$ and $BIC_\mathrm{quad} = 42.0$, therefore we adopted the linear fit as a final solution. The obtained linear ephemeris is presented in Table \ref{keltslineareph}. Taking the precise \textit{CHEOPS} and \textit{TESS} observations into consideration, we can exclude periodic TTVs with a semi-amplitude of about 3 min ($3\sigma$ upper limit), which is an improvement by a factor of 2 in comparison with the discovery O-C values. In the case of KELT-21b there are observations from three seasons, as well, which cover a time-baseline of about 2000 days. The relative scatter of the \textit{CHEOPS} and \textit{TESS} data is larger in comparison with the previous case. On the other hand, the newly obtained O-C data are more precise than the discovery data, see Fig. \ref{kelt21ttv}. The quadratic trend is not significant, i.e., there is no significant difference between the Bayesian Information Criterion of the linear and the quadratic fit ($BIC_\mathrm{lin} = 54.6$ and $BIC_\mathrm{quad} = 53.6$, respectively). Since no statistical justification for the quadratic fit, we can adopt the linear fit as a final solution. The obtained linear ephemeris is presented in Table \ref{keltslineareph}. Based on the \textit{CHEOPS} and \textit{TESS} O-C data we did not find periodic TTVs with a semi-amplitude larger than 3 min ($1\sigma$ upper limit). Since the relatively larger scatter in the new datasets, more and precise observations are needed to improve this value. In summary, we did not find any convincing evidence for an additional object in these systems, but via this procedure we improved further the orbital period of the planets (see Table \ref{keltslineareph} and compare the $P_\mathrm{orb}$ values to those of presented in Table \ref{photometryres}). Therefore, we can consider the $P_\mathrm{orb}$ parameter values presented in Table \ref{keltslineareph} as the final solutions.

\section{Conclusions}
\label{concl}

Using precise \textit{CHEOPS} and \textit{TESS} photometric observations, complemented with target spectroscopy, we analyzed three rapidly rotating planetary systems, i.e., KELT-17, KELT-19A, and KELT-21 from several viewpoints. We obtained new spectroscopic observations, which we used to derive stellar parameters of the planet hosts. Since the high effective temperature and the rapid rotation of the stars, the spectroscopic modeling was challenging in these cases and the resulting stellar parameters are not so precise as we expected before. On the other hand, based on the \textit{CHEOPS} photometric observations we were able to derive significantly improved system parameters in comparison with the previously published values. Based on these results we can conclude that KELT-17b and KELT-19Ab have smaller planet radius as found before, but in the case of KELT-17b this could be also due to the parameter degeneracy. For KELT-21b we could confirm the previously obtained system and planet parameters within $3\sigma$. The \textit{CHEOPS} light curves were also analyzed from the viewpoint of spin-orbit misalignment. Here we were able to confirm only that the gravity-darkening effect is very low in these cases. \textit{CHEOPS} data are too noisy to draw any conclusions on spin-orbit misalignment from the photometry alone. In addition, based on these analyses we can report on a tentative indication that the stellar inclination of KELT-21 is $I_* \approx 60$ deg.  

The \textit{CHEOPS} photometric observations, complemented with the available \textit{TESS} data were also used to search for transit duration variations and transit time variations in the systems. The search for long-term TDVs in the systems was motivated by the Kepler-13A planetary system, where orbital precession was identified, causing a long-term trend in the transit duration. In the cases of KELT-17b and KELT-19Ab we were able to exclude long-term TDVs causing orbital precession. The shorter transit duration of KELT-19Ab compared to the discovery paper is probably a consequence of a smaller planet radius. In the case of KELT-21b, there is an indication that a long-term TDV may exist in a connection with orbital precession, therefore this systems could be interesting from this viewpoint. More high-quality data are needed in the future to confirm, or reject the orbital precession/long-term TDV. Furthermore, via observed-minus-calculated diagrams of mid-transit times we probed the photometry data from the viewpoint of additional objects in the systems, but we did not find any convincing evidence. Based on the \textit{CHEOPS} and \textit{TESS} observations we set new upper limits on possible TTV semi-aplitudes and we were able to improve further the orbital period of the planets.


\section*{Acknowledgements}

We thank the anonymous reviewer for the helpful comments and suggestions. We also thank Dr. K. G. Isaak, the ESA CHEOPS Project Scientist, responsible for the ESA CHEOPS Guest Observers Programme, for the helpful discussions and support. This work was supported by the Hungarian National Research, Development and Innovation Office (NKFIH) grant K-125015, the PRODEX Experiment Agreement No. 4000137122 between the ELTE University and the European Space Agency (ESA-D/SCI-LE-2021-0025), the City of Szombathely under agreement No. 67.177-21/2016, and by the VEGA grant of the Slovak Academy of Sciences No. 2/0031/22. TP acknowledges support from the Slovak Research and Development Agency -- the contract No. APVV-20-0148. AC acknowledges financial support from the State Agency for Research of the Spanish MCIU through the ''Center of Excellence Severo Ochoa'' award for the Instituto de Astrophysics of Andalusia (SEV-2017-0709). CHEOPS is an ESA mission in partnership with Switzerland with important contributions to the payload and the ground segment from Austria, Belgium, France, Germany, Hungary, Italy, Portugal, Spain, Sweden, and the United Kingdom. The authors acknowledge the observing time awarded within the CHEOPS Guest Observers Programme No. 1 (AO-1) and the support from the Science Operations Centre. This paper includes data collected with the TESS mission, obtained from the MAST data archive at the Space Telescope Science Institute (STScI). Funding for the TESS mission is provided by the NASA Explorer Program. STScI is operated by the Association of Universities for Research in Astronomy, Inc., under NASA contract NAS 5-26555. This work has made use of data from the European Space Agency (ESA) mission Gaia (\url{https://www.cosmos.esa.int/gaia}), processed by the Gaia Data Processing and Analysis Consortium (DPAC, \url{https://www.cosmos.esa.int/web/gaia/dpac/consortium}). Funding for the DPAC has been provided by national institutions, in particular the institutions participating in the Gaia Multilateral Agreement.

\section*{Data availability}

The data underlying this article will be shared on reasonable request to the corresponding author. The reduced light curves presented in this work will be made available at the
CDS (\url{http://cdsarc.u-strasbg.fr/}).




\bibliographystyle{mnras}
\bibliography{Yourfile} 






\bsp	
\label{lastpage}
\end{document}